\documentclass[iop,apj,twocolumnappendix]{emulateapj}
\usepackage{amsmath,amssymb,amstext}
\usepackage{longtable, graphicx, subfigure,amsmath} 
\usepackage{lmodern}
\usepackage[T1]{fontenc}
\usepackage[breaklinks,colorlinks,citecolor=blue,linkcolor=magenta]{hyperref}

\usepackage{aas_macros}
\usepackage{natbib}
\newcommand{\Msun}{\ifmmode {M_{\odot}}\else${M_{\odot}}$\fi}
\newcommand{\Rsun}{\ifmmode {R_{\odot}}\else${R_{\odot}}$\fi}
\newcommand{\Lsun}{\ifmmode {L_{\odot}}\else${L_{\odot}}$\fi}
\newcommand{\lapprox }{{\lower0.8ex\hbox{$\buildrel <\over\sim$}}}
\newcommand{\gapprox }{{\lower0.8ex\hbox{$\buildrel >\over\sim$}}}
\def\amin{\ifmmode^{\prime}\else$^{\prime}$\fi}
\def\asec{\ifmmode^{\prime\prime}\else$^{\prime\prime}$\fi}

\newcommand{\degree}{\ifmmode {^\circ}\else$^\circ$\fi}
\newcommand{\halpha}{$H\alpha$\ }
\newcommand{\Ro}{\ifmmode {R_o}\else$R_o\ $\fi}
\newcommand{\lha}{\ifmmode {L_{H\alpha}/L_{bol}}\else$L_{H\alpha}/L_{bol}$\ \fi}

\shorttitle{Hyades Rotation Periods from {\it K2}}
\shortauthors{Douglas et al.}

\begin{document}


\title{{\it K2} Rotation Periods for low-mass Hyads and the Implications for Gyrochronology}
\author{S.~T.~Douglas\altaffilmark{1},
M.~A.~Ag{\" u}eros\altaffilmark{1},
K.~R.~Covey\altaffilmark{2}, 
P.~A.~Cargile\altaffilmark{3}, 
T.~Barclay\altaffilmark{4},
A.~Cody\altaffilmark{4,5},
S.~B.~Howell\altaffilmark{4},
T.~Kopytova\altaffilmark{6,7}
}
\altaffiltext{1}{Columbia University, Department of Astronomy, 550 West 120th Street, New York, NY 10027} 
\altaffiltext{2}{Western Washington University, Department of Physics \& Astronomy, Bellingham, WA 98225}
\altaffiltext{3}{Harvard-Smithsonian Center for Astrophysics, 60 Garden Street, Cambridge, MA 02138}
\altaffiltext{4}{NASA Ames Research Center, Moffett Field, CA 94035}
\altaffiltext{5}{NASA Postdoctoral Program Fellow}
\altaffiltext{6}{Max-Planck-Institut f\"ur Astronomie, K\"onigstuhl 17, 69117 Heidelberg, Germany}
\altaffiltext{7}{International Max-Planck Research School for Astronomy and Cosmic Physics at the University of Heidelberg, IMPRS-HD, Germany}

\begin{abstract}

As the closest open cluster to the Sun, the Hyades is an important benchmark for many stellar properties, but its members are also scattered widely over the sky. Previous studies of stellar rotation in the Hyades relied on targeted observations of single stars or data from shallower all-sky variability surveys. The re-purposed {\it Kepler} mission, {\it K2}, is the first opportunity to measure rotation periods ($P_{rot}$) for many Hyads simultaneously while also being sensitive to fully convective M dwarf members. 
We analyze {\it K2} data for 65 Hyads and present $P_{rot}$ values for 48. Thirty{\bf -}seven of these are new measurements, including the first $P_{rot}$ measurements for fully convective Hyads. For nine of the 11 stars with $P_{rot}$ in the literature and this work, the measurements are consistent; we attribute the two discrepant cases to spot evolution.
Nearly all stars with masses $\lapprox0.3$~\Msun\ are rapidly rotating, indicating a change in rotation properties at the boundary to full convection. When confirmed and candidate binaries are removed from the mass-period plane, only three rapid rotators with masses $\gapprox0.3$~\Msun\ remain. 
This is in contrast to previous results showing that the single-valued mass-period sequence for $\approx$600~Myr-old stars ends at $\approx$0.65~\Msun\ when binaries are included. 
We also find that models of rotational evolution predict faster
rotation than is actually observed at $\approx$600 Myrs for stars $\lapprox$0.9~\Msun.
The dearth of single rapid rotators more massive than $\approx$0.3~\Msun\ indicates that magnetic braking is more efficient than previously thought, and that age-rotation studies must account for multiplicity.

\end{abstract}

\keywords{stars:~evolution -- stars:~late-type -- stars:~rotation}
\maketitle

\section{Introduction}
As the nearest open cluster to the Sun \citep[$\approx$47~pc;][]{vanleeuwen2009}, the Hyades is a benchmark for understanding stellar properties. Many brighter Hyads were targeted by {\it Hipparcos}  \citep{hip, perryman1998}, allowing for precise astrometric and absolute photometric calibration. The cluster has also been surveyed for binary stars \citep[e.g.,][]{griffin1988,patience1998,reid2000,duchene2013} and for planets \citep[e.g.,][]{cochran2002,paulson2004,quinn2014}. Hyads have been used to test stellar models \citep[e.g.,][]{castellani2001} and to calibrate gyrochronology and metallicity scales \citep[e.g.,][]{skumanich72, barnes2003, paulson2003, delorme2011}.

The Hyades's age also contributes to its benchmark status: at 625$\pm$50~Myr \citep{perryman1998}, it is the oldest nearby open cluster.\footnote{The next nearest, older cluster is the $\approx$3-Gyr-old Ruprecht 147. At $D\approx300$~pc and in the direction of the Galactic Bulge, it is far harder to study \citep[][]{curtis2013}.}
The cluster's canonical age, however, has been called into question by \citet{brandt2015-1, brandt2015}, who fit rotating stellar models to main-sequence turnoff Hyads and find that the cluster is 750$\pm$100~Myr.\footnote{However, \citet{brandt2015-1, brandt2015} mention several potential sources of $\approx$100~Myr systematic uncertainties, and these are not included in their final quoted uncertainties.}
If the Hyades is older than previously thought, then the stellar evolution scales that it anchors, including gyrochronology, will need to be re-calibrated.

The Hyades's proximity aids observations of individual members, but presents challenges for studying the population as a whole. The $\approx$750 cluster members are scattered widely across the sky, covering $\gapprox$3600~deg$^2$. Ground-based observations designed to measure rotation periods ($P_{rot}$) typically have small fields-of-view ($\lapprox$1~deg$^2$), which would prevent them from obtaining light curves for more than one or two Hyads at a time. Previous campaigns to measure $P_{rot}$ in the Hyades have therefore been limited to observations of individual targets \citep{radick1987, radick1995, prosser1995} or to shallower large-area surveys designed to discover transiting exoplanets or other transients \citep[e.g., SuperWASP and HATNet;][]
{delorme2011, hartman2011}. These surveys have generally focused on solar-type stars ($0.7-1.2\ \Msun$), with almost no sensitivity to fully convective stars ($\lapprox$0.3~\Msun). 


After the second of its four reaction wheels failed, the {\it Kepler} spacecraft could no longer maintain pointing toward its original field. However, by observing in the ecliptic, stable pointing can be maintained using the two functioning reaction wheels to balance the torque on the spacecraft due to sunlight. The {\it K2} mission \citep{howell2014} stares at fields of view along the ecliptic, observing each for approximately 80 days.
The Hyades was targeted in {\it K2}'s Campaign 4, and the spacecraft's $\approx$100~deg$^2$ field-of-view provided an unprecedented opportunity to measure $P_{rot}$ for a large number of Hyads. Light curves were obtained simultaneously for 65 Hyads, and {\it Kepler}'s sensitivity enabled the first assembly of light curves for fully convective Hyads.


We present the results of our analysis of these {\it K2} data, including the first measurements of $P_{rot}$ for fully convective Hyads.\footnote{\citet{hartman2011} assign a \cite{prosser1995} rotator $M_*\approx0.22\ \Msun$, but we find $M_*=0.36\ \Msun$, and the star is a known visual binary \citep[WDS J04288+1617B;][]{mason2001}.} We describe the existing data used in our study, including membership catalogs, photometry, archival $P_{rot}$, and studies of multiplicity, in Section~\ref{data}. We present our mass measurements in Section~\ref{masses} and our {\it K2} data analysis in Section~\ref{k2}. Our results are in Section~\ref{res}, and we discuss their potential implications for gyrochronology in Section~\ref{disc}. We conclude in Section~\ref{concl}. 

\section{Archival Data}\label{data}
\subsection{Membership}\label{cats}
We draw Hyades members primarily from the \citet{goldman2013} catalog, which is an updated version of the \citet{roser2011} catalog used in \citet[][hereafter D14]{douglas2014}. \citet{goldman2013} and \cite{roser2011} identified candidate Hyads via the convergent point method and confirmed their membership using photometry: \citet{roser2011} combined $JHK$ data from the Two Micron All-Sky Survey \citep[2MASS;][]{2mass} with $r'$ photometry from the Carlsberg Meridian Catalogue 14 \citep[CMC14;][]{cmc14}. 

\citet{goldman2013} added $ugriz$ photometry from the Sloan Digital Sky Survey (SDSS) Data Release 8 \citep[][]{DR8paper}, $g_{P1}r_{P1}i_{P1}z_{P1}y_{P1}$ photometry from Pan-STARRS 1 \citep[][]{panstarrs, tonry2012}, and data from all four bands from the Wide-field Infrared Survey Explorer \citep[WISE;][]{wise}. The Pan-STARRS 1 and WISE data allow these authors to extend the cluster's membership to the stellar/sub-stellar boundary. \citet{goldman2013} also reject 13 stars that were identified as Hyads by \citet{roser2011} and add 62 low-mass stars and brown dwarfs. The final \citet{goldman2013} catalog contains 773 members. 

\citet{roser2011} estimate contamination to their catalog based on distance from the cluster center ($d_c$): it is $1\%$ for stars with $d_c\le9$~pc, $7.5\%$ for $9 <d_c\le18$~pc, and $30\%$ for $18 <d_c\le30$~pc. In D14, we converted these percentages to membership probabilities, $P_{mem}$, by subtracting the contamination percentage from $100\%$.  

\citet{goldman2013} simulate field star contamination as a function of $d_c$ and absolute $K$ magnitudes ($M_K$) but do not give their full results, stating only that the contamination is $<$10\% for $d_c\le18$~pc, $\le$17\% for stars with $18 <d_c\le30$~pc and $7.5 < M_K < 9.5$~mag, 
and $\le$30\% for $d_c\le30$~pc and $M_K=9$. 
These contamination rates appear consistent with those from \citet{roser2011}. Therefore, we also calculate distance-based $P_{mem}$ for the new \citet{goldman2013} members using the method described in D14. 

We supplement the \citet{goldman2013} catalog with new Hyades members found by our analysis of All Sky Automated Survey \citep[ASAS;][]{ASAS} data (Cargile et al., in prep.).
We follow the method outlined by \citet{vanleeuwen2009} and consider stars within 26\degree\ and 20 pc of the cluster center. We identify 170 cluster members with reduced proper motions  ($\mu$) satisfying $-170<\mu_{\parallel}<-60$ and $-20<\mu_{\perp}<20$~mas~yr$^{-1}$ and distances 
distances obtained by {\it Hipparcos} \citep{hip}. 
We add stars from the \citet{radick1987} and \citet{delorme2011} catalogs that were too bright or too faint to be included in the {\it Hipparcos} survey, so that this catalog includes 209 stars, all of which are listed in Table 6 of D14.
All but 13 of the Hyads identified in this manner were also identified by \citet{roser2011} and \citet{goldman2013}. 
We add these 13 stars to our final Hyades catalog, resulting in a membership catalog of 786 stars with $P_{mem} \ge 70\%$. 


\begin{figure}[!t]
\centerline{\includegraphics[width=\columnwidth]{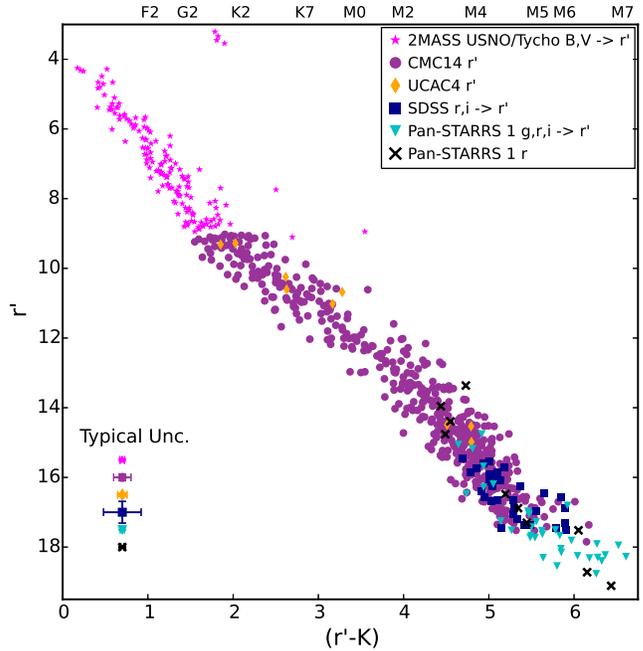}}
\caption{CMD for the Hyads in our final catalog.
The symbols represent different sources of photometry, and whether conversions were necessary to obtain the $r'$ magnitudes. We could not convert $r_{P1}$ into $r'$ for a subset of the \citet{goldman2013} stars (see Section~\ref{photometry}), so we retain the $r_{P1}$ magnitudes and mark these stars with crosses. The typical uncertainty (propagated from the source photometry) is shown in the lower left.}
\label{fig:rsource} 
\end{figure}

\begin{deluxetable*}{llcccc}[t]
\tablewidth{0pt}
\tabletypesize{\scriptsize}
\tablecaption{Hyads with multiple $P_{rot}$ measurements \label{tab:protdiff}}
\tablehead{
\colhead{Roser}& \colhead{Object}& \colhead{\citet[][]{prosser1995}}& \colhead{\citet{delorme2011}}& \colhead{ASAS-derived}& \colhead{{\it K2}-derived}\\ 
 \colhead{No.}& \colhead{Name}& \colhead{$P_{rot}$ (d)}& \colhead{$P_{rot}$ (d)}& \colhead{$P_{rot}$ (d)\tablenotemark{a}}& \colhead{$P_{rot}$ (d)}
}
\startdata
68\tablenotemark{b} & HIP 16908& \nodata& 10.57& 11.17& \nodata\\
78 & EPIC 210865372& \nodata& 12.30& \nodata& 12.32\\
85\tablenotemark{b} & EPIC 211058178& \nodata& \nodata& 8.37& 12.57\\
125\tablenotemark{b} & EPIC 210772796& \nodata& 11.45& \nodata& 11.60\\
133 & EPIC 210771915& \nodata& 13.51& 7.19\tablenotemark{c}& 13.85\\
140\tablenotemark{d} & HIP 19207& \nodata& 11.98& 14.03& \nodata\\
144\tablenotemark{b} & HIP 19263& \nodata& 12.30& 11.60& \nodata\\
148 & EPIC 211049709& \nodata& 9.35& \nodata& 9.39\\
202 & EPIC 210744818& \nodata& 12.95& \nodata& 12.84\\
203 & EPIC 210470484& \nodata& 2.35& \nodata& 2.38\\
267\tablenotemark{b} & EPIC 210786154& \nodata& 9.90& 9.96& 10.02\\
288 & EPIC 210674406& \nodata& 11.60& 11.87& 11.47\\
303\tablenotemark{b} & EPIC 210741091& \nodata& 10.84& \nodata& 10.97\\
339\tablenotemark{b} & 2M J0427+1415& \nodata& 12.78& 12.81& \nodata\\
345\tablenotemark{b} & HIP 20827& \nodata& 9.70& 10.28& \nodata\\
355\tablenotemark{b} & EPIC 210651981& 2.42& 2.42& \nodata& 2.44\\
428 & HIP 21256& \nodata& 12.69& 13.59& \nodata\\
438\tablenotemark{b} & 2M J0434+1133& \nodata& 11.03& 11.59& \nodata\\
486 & HIP 21723& \nodata& 10.85& 11.30& \nodata\\
514\tablenotemark{b} & 2M J0443+1704& \nodata& 10.31& 10.18& \nodata\\
558\tablenotemark{b} & HIP 22350& \nodata& 9.69& 9.92& \nodata\\
564\tablenotemark{b} & HIP 22394& \nodata& 6.90& 7.07& \nodata\enddata
\tablecomments{
EPIC 210359769 and EPIC 210675409 also have $P_{rot}$ measurements from \citet{delorme2011} and were targeted by {\it K2}, but we were unable to measure a new $P_{rot}$ for either (see Table~\ref{tab:k2bad}).  There was no overlap between the listed samples and those from \citet{radick1987, radick1995} and \citet{hartman2011}
}
\tablenotetext{a}{Cargile et al.~(in prep)}
\tablenotetext{b}{Confirmed binary}
\tablenotetext{c}{Half-period harmonic}
\tablenotetext{d}{Planet host; $P_{orb}\approx6.09$~d \citep{quinn2014}. 
In D14 we chose to use the ASAS-derived $P_{rot}$ value for this star, and for consistency we use that value here as well.}
\end{deluxetable*}
\subsection{Photometry}\label{photometry}

As in D14, we use $(r'-K)$ as our primary proxy for stellar temperature.  
Optical-near infrared (NIR) colors allow a broader dynamic range than is possible with a narrower color index, particularly for lower-mass stars. For example, when using NIR photometry alone, M dwarfs only show colors $0.9\ \lapprox\ (J-K)\ \lapprox\ 1.2$ mag, but stars in this same mass range are spread across $3.3<(r'-K)<8.0$. While nearly all the stars in our catalog have 2MASS $K$-band magnitudes, the large range in $r'$ magnitudes ($\approx$15 mag) for these objects meant that we had to obtain this photometry from multiple sources. 

CMC14 includes $r'$ measurements for $\approx$$10^8$ stars with declinations between $-30$\degree\ and $50$\degree\ and $9\ < r'\ \lapprox\ 17$ mag. We use this photometry for 522 Hyads falling within this magnitude range; for the members identified by \cite{roser2011}, we use the CMC14 magnitudes and errors listed in that catalog. For 17 Hyads with $10\ \lapprox\ r'\ \lapprox\ 14$~mag that do not appear in CMC14, we use $r'$ magnitudes from the 4$^{\rm th}$ U.S.~Naval Observatory CCD Astrograph Catalog \citep[UCAC4;][]{zacharias2012}, which includes $g'r'i'$ magnitudes from APASS \citep{apass}. In total, 539 Hyads have $r'$ data from CMC14 or UCAC4. 

The remaining Hyads are too bright or too faint to appear in CMC14 or UCAC4. We convert photometry from other systems into $r'$ magnitudes for these stars.

$\bullet$ Thirty-three new \citet{goldman2013} members have SDSS $ri$ photometry.  We convert those magnitudes to $r'$ using transformation equations from \citet{jester2005} and documented online.\footnote{\url{http://www.sdss.org/dr7/algorithms/jeg\_photometric\_eq\_dr1.html}}


$\bullet$ For 44 new \citet{goldman2013} members with photometry in all three Pan-STARRS 1 $g_{P1}r_{P1}i_{P1}$ bands, we use appropriate equations from \citet{tonry2012} and \citet{jester2005} to convert these magnitudes to SDSS $r$ and $i$ magnitudes, and from there to $r'$.

$\bullet$ Eighteen new members identified by \citet{goldman2013} were detected in only one or two Pan-STARRS 1 bands, and lack the photometry needed for a conversion to $r'$.  The seven stars with detections only in $g_{P1}$ or $i_{P1}$ do not have previous $P_{rot}$ measurements, nor are they {\it K2} targets, so the lack of $r'$ photometry does not impact our analysis. Eleven stars were only detected in $r_{P1}$; this includes one {\it K2} target, EPIC 210489654. We retain the unconverted $r_{P1}$ magnitude for these stars. 

$\bullet$ For 152 bright stars with $r<9$ mag, we use the \citet[][]{jester2005} relations to convert the USNO-A2.0 and Tycho 2 Johnson $B$ and $V$ magnitudes included in the 2MASS catalog to SDSS $r$ magnitudes. Fortuitously, these stars fall into the appropriate color range for which the \citet{bilir2008} transformation from 2MASS colors to $(r-i)$ can be used. Finally, we use these $r$ and $(r-i)$ values to obtain $r'$ by applying the \citet{jester2005} relation, as above. 

The typical uncertainty for these $r'$ magnitudes depends on the source catalog. After applying the conversions discussed above to 2MASS and SDSS photometry, the uncertainties are generally $\lesssim$0.1 and $\approx$0.3 mag, respectively. For CMC14 photometry, the uncertainty is $\approx$0.1 mag; for UCAC4, $\approx$0.05~mag. The sources for all of our photometry are shown in the color-magnitude diagram (CMD) for the cluster presented in Figure~\ref{fig:rsource}.


\subsection{Archival Rotation Periods}\label{prot}

In D14 we assembled $P_{rot}$ measurements for Hyads from \citet{radick1987, radick1995};  \citet{delorme2011}; and from an analysis of ASAS data (Cargile et al., in prep) 
into a catalog of 87 rotators. We now restore eight additional stars with measured $P_{rot}$ that are confirmed binaries, bringing the total number of known rotators from these studies to 95. 

We then add $P_{rot}$ measured by \citet{hartman2011} for stars in our catalog of likely members. These authors used light curves from the HATNet survey to search for photometric variability among nearby field K and M dwarfs. \citet{hartman2011} present $P_{rot}$ measured at two different stages in their analysis: after applying external parameter decorrelation and then after applying a trend filtering algorithm. We required that the quality flags on both periods be 0 and that the periods \citet{hartman2011} measured at both stages agree to within one day. Although nine \citet{hartman2011} targets match objects in our Hyades catalog, two had bad quality flags, and another showed different periods after the two different filtering algorithms were applied; we add the six remaining $P_{rot}$ to our catalog

Finally, we add two rotation periods from \citet{prosser1995}. These authors observed stars from several open clusters and searched for periodic variability. They targeted three Hyads, and found two to be variable. 

\begin{figure}[t]
\centerline{\includegraphics[width=\columnwidth]{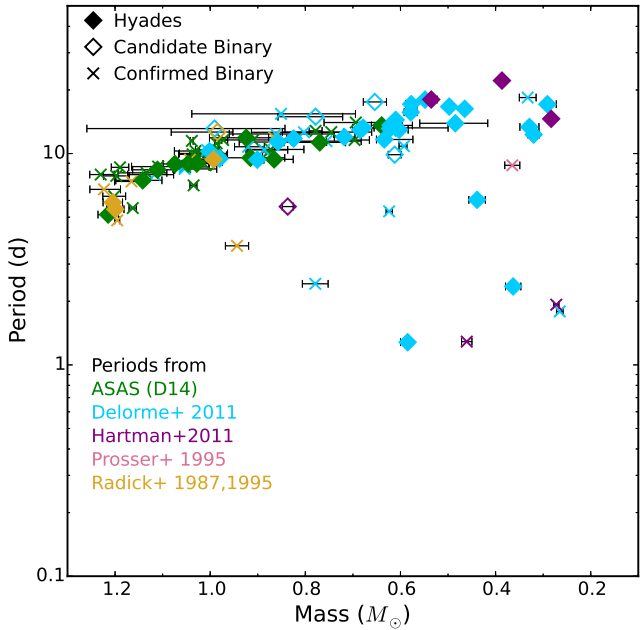}}
\caption{Mass-period distribution for Hyads with period measurements in the literature. The color indicates the source of the $P_{rot}$; empty diamonds and crosses denote photometrically identified candidate binaries and confirmed binaries, respectively. We also include the mass uncertainties; in general, $\sigma_{M*}\approx 2-3\%$, which is smaller than or comparable to the symbol size. Distance uncertainties dominate the $M_*$ uncertainties; the stars with large $\sigma_{M_*}$ have $\sigma_D\approx10-20$~pc. The error bars only represent systematic uncertainties from our mass calculation, and do not reflect, e.g., systematics in the model or excess $K$-band flux due to an unresolved companion.}
\label{fig:periodmass} 
\end{figure}

In total, we have 102 Hyads with $P_{rot}$ measurements from the literature. However, 48 are confirmed binaries, and eight more are candidate binaries (see discussion below). This leaves 44 single\footnote{Because not all of these have been surveyed for binarity, this is an upper limit on the true number of single stars in this sample.} Hyads with measured $P_{rot}$ before the addition of the {\it K2} data. The mass-period relationship for these 102 Hyads is shown in Figure~\ref{fig:periodmass}.

\begin{figure}[t]
\centerline{\includegraphics[width=\columnwidth]{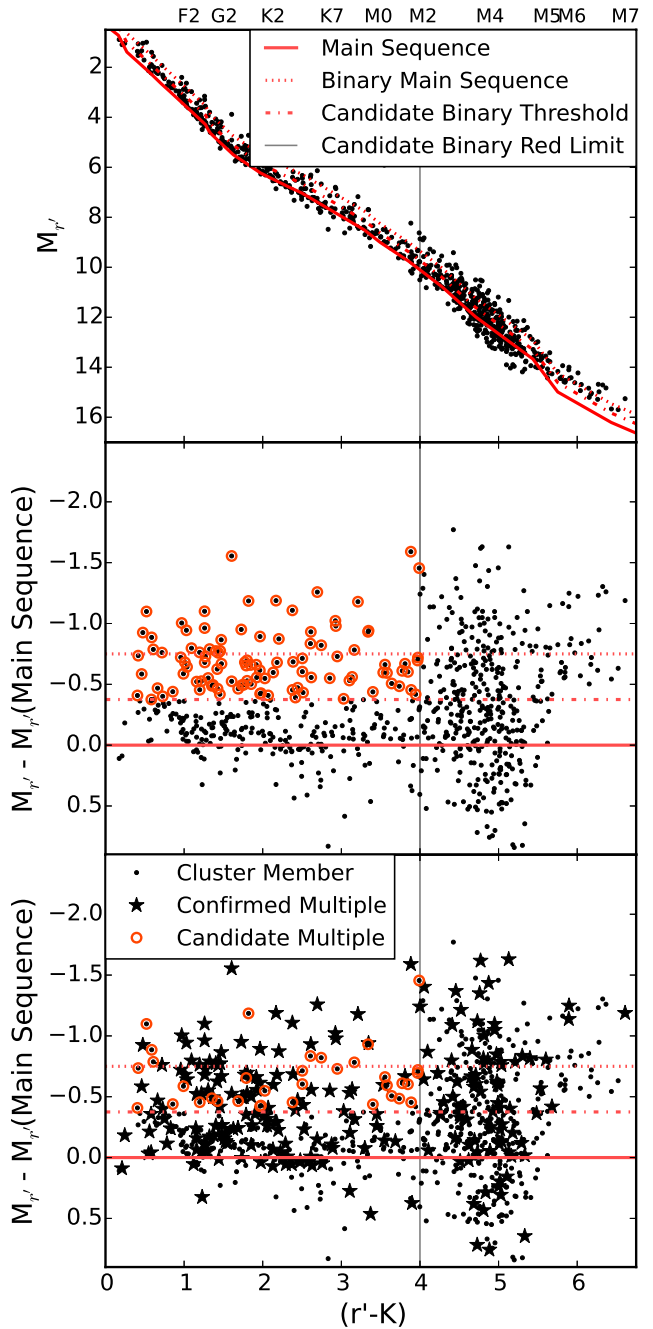}}
\caption{{\it Top} --- CMD of the Hyades. The solid line traces out the main sequence, identified using the SEDs assembled by \citet{adam2007},
and the dotted line the binary main sequence. 
We label any star  lying above the dot-dashed line halfway between these two sequences as a candidate binary \citep[as in][]{hodgkin1999}. Stars with $(r'-K)\ \gapprox\ 4$ have a wider range of magnitudes at a given color, and do not show an easily identified main sequence, so we do not search for photometric binaries in this color range.
{\it Middle} --- Residuals between each star's $M_{r'}$ and the model main-sequence magnitude. Photometrically identified candidate binaries are shown as orange circles. This method is primarily sensitive to $\approx$equal-mass binaries.
{\it Bottom} --- Same as above, with confirmed binaries from the literature shown as black stars (identifications are only complete for stars with measured $P_{rot}$). Confirmed binaries can be found at all distances from the main sequence, illustrating the limitations of this approach to binary identification.}
\label{fig:binarycmd} 
\end{figure} 

\begin{deluxetable*}{lcccccp{6cm}}[t]
\tablewidth{0pt}
\tabletypesize{\footnotesize}
\tablecaption{Confirmed and candidate binary and planetary systems among the {\it K2} targets and Hyads with measured $P_{rot}$ \label{tab:bin} }
\tablehead{
\colhead{Roser} & \colhead{HIP} & \colhead{2MASS J} & \colhead{EPIC} & \colhead{Cand?\tablenotemark{a}} & \colhead{$P_{orb}$ (d)} & \colhead{Source\tablenotemark{b}} 
}
\startdata
\multicolumn{7}{c}{Binaries}\\
\tableline                  
207	&	\nodata	&	04174767+1339422	&	210408563	&	N	&	\nodata	&	\citet{guenther2005}	\\ 
214	&	\nodata	&	04181077+2317048	&	211019716	&	Y	&	1.88	&	\citet{griffin1982, pourbaix2004, kopytova2016}	\\ 
216	&	\nodata	&	04181926+1605181	&	\nodata	&	Y	&	\nodata	&		\nodata							\\ %
267	&	20482	&	04232283+1939312	&	210786154	&	Y	&	\nodata	&	\citet{morzinski2011}	\\ 
270	&	20485	&	04232526+1545474	&	\nodata	&	Y	&	\nodata	&	\citet{morzinski2011}	\\ 
275	&	\nodata	&	04235070+0912193	&	\nodata	&	N	&	5.26	&	\citet{griffin2012}	\\ 
284	&	20553	&	04241244+1445295	&	\nodata	&	Y	&	\nodata	&	\citet{patience1998, kopytova2016}	\\ 
293	&	20577	&	04242831+1653103	&	\nodata	&	Y	&	\nodata	&	\citet{patience1998, kopytova2016}	\\ 
303	&	\nodata	&	04251456+1858250	&	210741091	&	N	&	\nodata	&	\citet{morzinski2011}	\\ 
\tableline                  
\multicolumn{7}{c}{Planets}\\
\tableline                  
140	&	19207	&	04070122+1520062	&	\nodata	&	N	&	6.09	&	\citet{quinn2014}\\	 
169  & \nodata &  04130560+1514520 &   210490365 & N & 3.48 & \citet{mann2015,david2016} 
\enddata
\tablenotetext{a}{Photometric candidate binary identified in Section \ref{binaries}.}
\tablenotetext{b}{If no source is given, we have identified the star as a photometric candidate binary, but its multiplicity is not confirmed.}
\tablecomments{This table is available in its entirety in a machine-readable form in the online journal. A portion is shown here for guidance regarding its form and content.}
\end{deluxetable*}
\subsection{Companions} \label{binaries}

Close companions impact the rotational evolution of stars, potentially biasing our analysis. Additional flux from a companion star could also contaminate our mass 
calculations. 
Therefore, we try to identify all confirmed or candidate binaries and remove these from our gyrochronology analysis. 

In D14, we used SIMBAD and \citet{delorme2011} to identify nine binaries in the Hyades. 
We now carry out a more thorough search of the literature for binaries among known Hyades rotators and {\it K2} targets. 

We begin with the catalog compiled by \citet{kopytova2016}. These authors combined the results of several multiplicity surveys \citep{patience1998, mason2001, mermilliod2009, morzinski2011, duchene2013}, to which they added their own AstraLux lucky imaging observations and visual companions identified in archival Hubble Space Telescope (HST) images. Of 724 \citet{roser2011} Hyads, \citet{kopytova2016} identify 213 as binary or multiple systems. 

We add confirmed binaries from \citet{barradoynavascues1996}, \citet{tokovinin1997}, \citet{reid1997}, \citet{paulson2004}, \citet{pourbaix2004}, \citet{guenther2005}, \citet{delorme2011}, \citet{griffin2012}, and \citet{janson2014}. Other known rotators and/or {\it K2} targets were  included in previous surveys, but these did not resolve any companions \citep[][]{gizis1995,patience1998,reid2000,mason2001,richichi2002,paulson2004,duchene2013}. 
There are also two confirmed planets in the Hyades; both have short orbital periods \citep{quinn2014,mann2015}.
In total, we find two planet hosts and 70 confirmed multiple systems among the known rotators and {\it K2} targets.

As in D14, we also identify candidate unresolved binaries that are overluminous for their color (see Figure~\ref{fig:binarycmd}). We use  model stellar SEDs assembled by \citet{adam2007} to identify the single-star main sequence, then identify a binary main sequence offset by 0.75~mag for a given color from that of single stars \citep[as in][]{steele1995}.  We then label stars with $(r'-K)<4$ as candidate binaries if they lie above the midpoint between the single-star and binary main sequences \citep{hodgkin1999}. This method is biased toward binaries with equal masses, so that we are certainly missing candidate binaries with lower mass ratios. Indeed, the bottom panel of Figure~\ref{fig:binarycmd} shows confirmed binaries at all distances from the main sequence. While further observations are required to confirm the binary status of all cluster members, this photometric approach does allow the identification of a significant number of binaries. 

We only apply this method to stars with $(r'-K)<4$ because the single-star main sequence is less apparent for stars redder than this value. 
The observed spread in magnitudes could be due to binary systems at a variety of mass ratios, or to uncertainties in distance or magnitude for these faint red stars. Identifying even candidate binaries in this regime therefore requires more information than just photometry.

In total, after our analysis of the {\it K2} data (Section \ref{k2}), we find that 63 Hyads with measured periods are confirmed binaries, two are confirmed planet hosts, and eight are candidate binaries. Another six {\it K2} targets for which we could not measure $P_{rot}$ are confirmed binaries, and one is a candidate. We list all confirmed and candidate binaries in Table~\ref{tab:bin}, and exclude them from our gyrochronology analysis in Section~\ref{disc}. 


\section{Stellar masses}\label{masses}
As in D14, we estimate stellar masses ($M_*$) by linearly interpolating between the $M_K$ and $M_*$ points given by \citet{adam2007}, who list 
$M_*$ and spectral energy distributions (SEDs) for B8-L0 stars. 

We calculate $M_K$ using {\it Hipparcos} parallaxes \citep{hip} where possible to determine distances ($D$) to individual stars. When {\it Hipparcos} parallaxes are not available, we use the secular parallaxes published by \citet{roser2011} or \citet{goldman2013}.  The 13 ASAS 
stars that are not in the \citet{roser2011} or \citet{goldman2013} catalogs do not have {\it Hipparcos} parallaxes, and for these stars we assume $D=47$~pc \citep{vanleeuwen2009} with an uncertainty $\sigma_D=\pm10$~pc. 

We calculate the $M_K$ uncertainties by propagating the $m_K$ and $D$ uncertainties for each star, and then determine the $M_*$ uncertainties by passing the $\pm1\sigma_{M_K}$ values through the same $M_*$ calculation as above. Most of the uncertainties are small ($\sigma_{M*}\approx2-3\%$), but in a few cases, significant distance uncertainties lead to $\sigma_{M*}\approx 20-30\%$. 
Figure~\ref{fig:periodmass} includes these $\sigma_{M}$ values.
These are simply the systematic uncertainties from our calculation, and do not reflect other sources of uncertainty, such as our choice of model or $K$-band excesses due to a binary companion.

\section{{\it K2} data}\label{k2}

\begin{figure}[t]
\centerline{\includegraphics[width=\columnwidth]{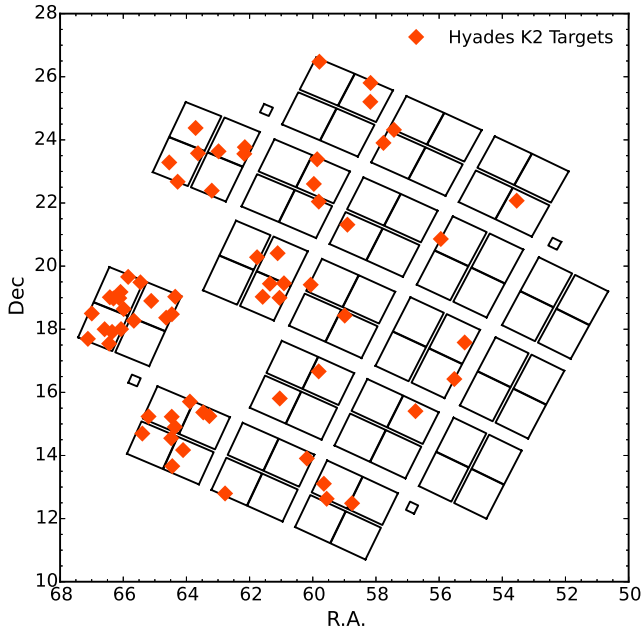}}
\caption{The {\it K2} Campaign 4 field of view, with our 65 Hyades targets shown as orange diamonds. Because the cluster is so large on the sky, many targets are near the edges of the field of view, and therefore have distorted PSFs. Two of the spacecraft's detector modules are no longer functioning. } 
\label{fig:k2fov} 
\end{figure} 

The pointing in {\it K2} is held in an unstable equilibrium against solar pressure by the two functioning reaction wheels. The spacecraft rolls about the boresight by up to 1 pixel at the edge of the focal plane. 
To correct for this, thrusters can be fired every 6 hours (if needed) to return the spacecraft to its original position. 
This drift causes stars to move on the focal plane in a diagonal pattern, inducing sawtooth-like signal in the 75-day light curve for each star (Van Cleve et al.~in prep.). 

For stars with intrinsic long-timescale, high-amplitude variability, the drift has minimal impact on measurements of $P_{rot}$. For rotational variability on short timescales ($\lapprox1$~d) or with low amplitudes, we must use caution to avoid removing or distorting the rotational signature while applying instrumental corrections to the light curves.

\begin{figure*}[t]
\centerline{\includegraphics[width=2\columnwidth]{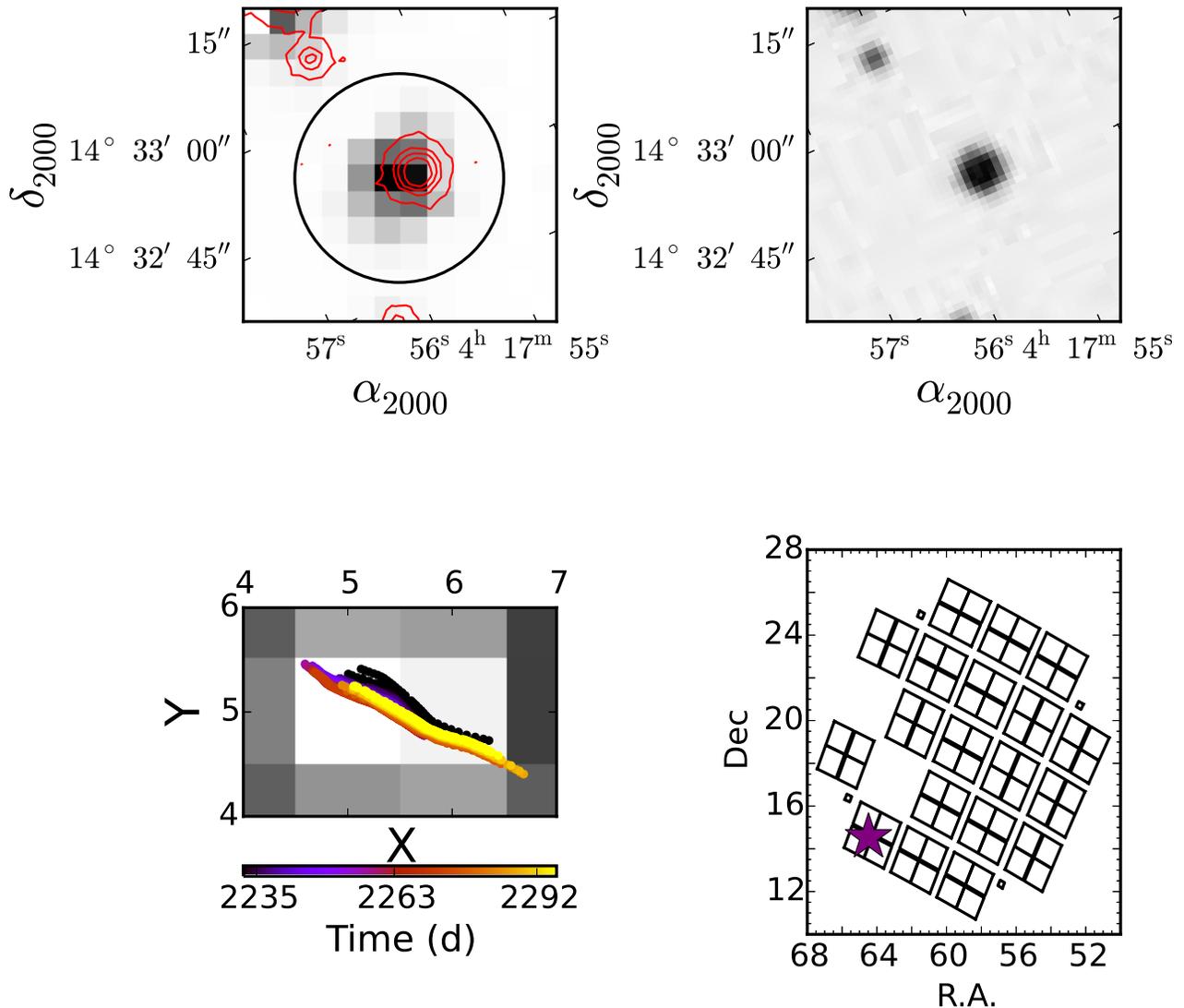}}
\caption{Diagnostic plots for EPIC 210451321; plots for all other {\it K2} Hyades targets are found in the electronic edition. {\it Top left} --- {\it K2} data in greyscale with the aperture selected in Section~\ref{k2prot} shown in black. The DSS2 red image of the target is overlaid as red contours. The WCS in the {\it K2} target pixel files does not perfectly match the WCS in the DSS images, particularly near the edges of the field of view where distortion is significant. {\it Top right} --- DSS2 red image of the target, rotated to match the orientation of the {\it K2} data. In this case, nearby companions are clearly visible within the {\it K2} pixel stamp, but they are excluded from the aperture used to extract the light curve. 
{\it Bottom left} --- {\it K2} data in greyscale (colors reversed from above), with the centroid drift overlaid. {\it Bottom right} --- The {\it K2} footprint for Campaign 4; the purple star indicates the position of the target.}
\label{fig:fourby} 
\end{figure*} 

\begin{figure*}[t]
\centerline{\includegraphics[width=2\columnwidth]{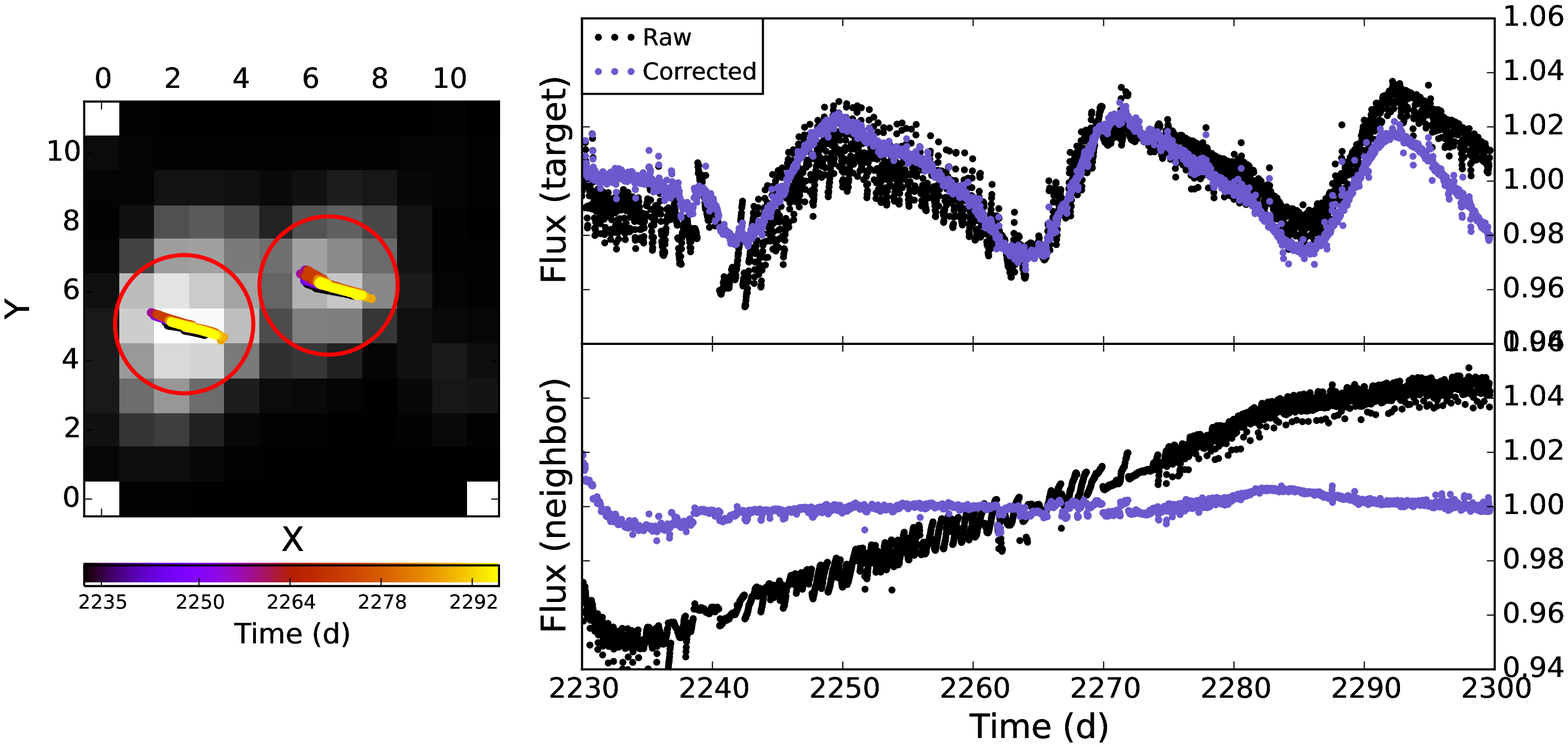}}
\caption{
{\it Left} --- Co-added {\it K2} images of EPIC 210736105 (center) and its nearby neighbor (left). We track the centroids with the {\it daofind} function in {\it photutils} and extract light curves through circular apertures with $r=2$~pixels (red). The centroid positions are colored by date; black points were taken at the beginning of the campaign and yellow at the end. 
{\it Right} --- {\it K2} light curves of EPIC 210736105 (top) and its nearby neighbor (bottom). The raw light curves are shown in black and corrected light curves in purple. Although the light curves are noisy, the periodic signature of the target clearly dominates any signal from the neighbor. Therefore, we simply restrict the centroid box and aperture size to extract a final light curve for EPIC 210736105.} 
\label{fig:neighbor} 
\end{figure*} 

We analyze {\it K2} Campaign 4 data for 65 Hyads identified in Section~\ref{cats} and with {\it Kepler} magnitudes $K_p>9$ mag (see Figure~\ref{fig:k2fov}). Sixty-two of these stars have $P_{mem}>90\%$ as calculated in Section \ref{cats}. We exclude an additional 14 Hyads with $K_p<9$, as they are saturated and therefore not amenable to our light-curve extraction method. The $K_p$ values in the EPIC catalog are only approximate---they are computed from survey photometry---so among the 65 Hyads are a few stars with $9\ \lapprox\ K_p\ \lapprox\ 10$ that also are too bright for our method (see Table \ref{tab:k2bad}). 

We describe how we extract light curves in Section \ref{lcextract}, how we account for nearby stars in Section \ref{neighbors}, and how we remove instrumental signatures in Section \ref{k2prot}.\footnote{Our python code for light curve extraction can be found at \url{https://github.com/stephtdouglas/k2phot/tree/hyades_paper/} and our python code for detrending/$P_{rot}$ measurement can be found at \url{https://github.com/stephtdouglas/k2spin/tree/hyades_paper/}.}

\subsection{Light Curve Extraction}\label{lcextract}

All 65 of our targets were observed in long-cadence, meaning exposures lasted 29.4~min. Eight of our targets were also observed in short-cadence (58.89~s exposures), but we restrict our analysis to the long-cadence data.

The {\it K2} data are provided as target pixel files, which include basic information about the target, individual exposure frames for each time point, pipeline-calculated background counts, and quality flags for each exposure. Quality flags indicate problems with the exposure, noting for example when data were obtained during the thruster fires. Before further analysis, we remove any exposures with quality flags not equal to 0. 

The Campaign 4 data processed by the {\it K2} pipeline is background-subtracted, but the pipeline background calculation is not reliable. The background is determined by fitting a low-order polynomial to the global background, which misses smaller local variations, especially those caused by dust near the Pleiades.\footnote{See \url{http://keplerscience.arc.nasa.gov/k2-data-release-notes.html\#k2-campaign-4}} 

We therefore perform our own background subtraction after adding the pipeline-calculated background flux back to the data. We compute the median flux of all pixels in each exposure, rejecting pixels identified as $3\sigma$ outliers over three iterations. After three iterations, the sigma-clipped median generally matches the levels in a visibly source-free region, and we take this value as the background level.

We calculate a flux-weighted centroid to determine the star's position in each exposure. We follow \citet{howell2006} in using a nine-by-nine pixel box around the nominal target position given by the exposure's header. 
We then place soft-edged circular apertures on the star's position that we calculated for each exposure, so that the aperture moves to follow the star. Tracking the motion of the star in this manner does not entirely remove variations in pixel sensitivity, but it does prevent nearby stars from entering the aperture. The soft-edged apertures include a fraction of the counts in pixels that are not entirely within the aperture. This fraction is equal to the fraction of the pixel covered by the aperture. 

We carry out all aperture photometry and some source detection using the {\it photutils} package.\footnote{\url{https://github.com/astropy/photutils}, Version 0.2} We extract light curves for each target using aperture radii $r$ ranging from $2\leq r\leq6.5$ pixels in 0.5~pixel increments. The optimal light curve for analysis is chosen at a later stage.

\subsection{Accounting for Nearby Stars}\label{neighbors}


We co-add the individual images for each star and search for any nearby sources automatically and by eye. These co-added images are included in our diagnostic plots for each target (Figure~\ref{fig:fourby}).

First, we use the {\it daofind} function from {\it photutils}\footnote{This is a Python implementation of the IRAF DAOFIND routine.} with a low threshold for detecting nearby stars. We set {\it sharphi}=5, {\it sharplo}=0.1, and {\it FWHM}=2.5. Because the flux-weighted method will simply find the brightest spot in the frame, we also extract light curves for all detected objects using {\it daofind} to track the centroids.

When the sources are well separated on the chip (i.e., the neighbor's centroid is $\gapprox$$2-3$ pixels), we did not find any significant variability in the neighbors' light curves.  Therefore, we simply restrict the maximum aperture radius for the primary target so that light from the neighbor does not affect further analysis of the target star. 

We also search by eye for any sources in the coadded image that were missed by {\it daofind}. All the neighboring stars are real objects that also appear on archival images, but are either too faint or too close to the edge of the pixel stamp to be detected by {\it daofind}. 
Our moving aperture method for constructing light curves requires that sources be automatically detected either by daofind or a flux-weighted centroid, so that the source's centroid position can be used to define the source's aperture in each exposure. We are therefore unable to produce light curves for faint neighbors that {\it daofind} cannot detect.  As above, in cases where we visually identify a faint neighbor, 
we restrict the maximum aperture radius for the primary target to exclude these neighbors wherever possible. 

We deviate from our standard extraction procedure in two cases where a bright neighbor on the chip biases the flux-weighted centroid of the target. For EPIC 210736105, we simply use a smaller five-by-five pixel box to calculate the flux-weighted centroid, which yields the correct position for the star. 
We also reduce the maximum aperture size to 2.0 pixels. The neighbor shows no evidence of periodic variability (Figure~\ref{fig:neighbor}).

\begin{figure*}[t]
\centerline{\includegraphics[width=2\columnwidth]{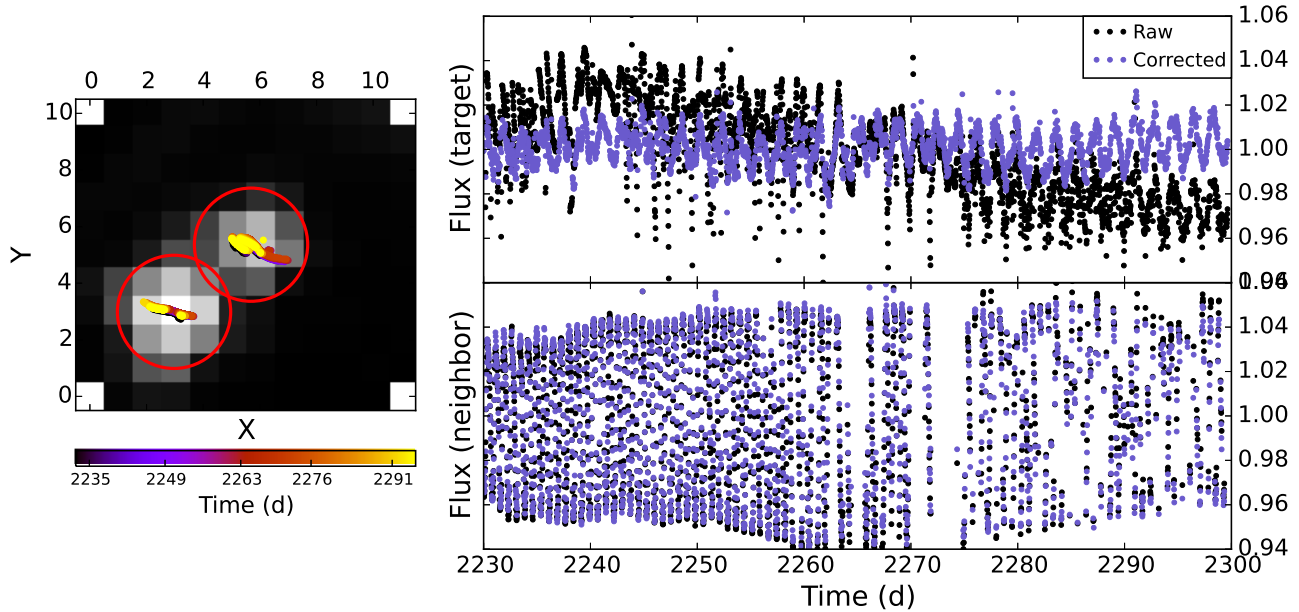}}
\caption{
Same as Figure \ref{fig:neighbor}, but for EPIC 210963067  and its nearby neighbor (at center and lower left, respectively, in postage stamp at left). In this case both the target and neighbor show periodic variability on different timescales. Furthermore, restricting the centroid box does not prevent the brighter neighbor from biasing the centroid. Therefore, we use the light curves shown above, which were extracted using {\it daofind} centroid, for our analysis in Section \ref{k2prot}.}
\label{fig:neighbor2} 
\end{figure*}

Restricting the centroid calculation box for EPIC 210963067 does not remove the bias towards the brighter neighbor, and the neighbor is also variable (Figure \ref{fig:neighbor2}). We therefore track the centroid of both the target and its neighbor using {\it daofind}. 
This is the only star for which we significantly change our light curve extraction technique based on a neighboring star. 

We also must account for the {\it K2} imager's large pixel scale (4\arcsec), which can produce blended point spread functions (PSFs) even when nearby objects are well-resolved in other surveys. 
We therefore compare the {\it K2} images by eye to Digital Sky Survey (DSS), SDSS (when available), and 2MASS images of our targets.\footnote{We use the IRSA/IPAC finder chart application to obtain these images: \url{http://irsa.ipac.caltech.edu/applications/finderchart/}.} A DSS or SDSS image of each target is shown in Figure~\ref{fig:fourby}, rotated into the frame of the {\it K2} image. In many cases these neighboring stars are too close to our target to be separated, and we flag these targets as having blended neighbors. 

Flags indicating whether a companion was identified by eye and whether it is blended in the target aperture are given in Table \ref{tab:k2}. A neighbor flag of ``Y'' indicates that a neighbor was identified. For these targets, a blended flag of ``Y'' indicates that the neighbor is definitely inside the {\it K2} aperture, ``M'' indicates that there is possible contamination or overlapping PSFs that can still be separated, and ``N'' indicates that the neighbor is well-separated. 

\subsection{Detrending Light Curves and Measuring $P_{rot}$}\label{k2prot}


As noted earlier, {\it K2} light curves contain instrumental signals that can hinder measurements of intrinsic stellar variability.  Given the 6-hour timescale for the dominant instrumental signature in {\it K2} light curves, however, disentangling intrinsic variability from instrumental effects can be difficult for stars with short-period, low-amplitude astrophysical signals. To decouple these signatures as best we can, we adopt an iterative approach that first removes intrinsic stellar signals from the light curve, so that instrumental effects can be isolated and corrected before further analyzing the stellar variability. 

We describe this iterative approach in detail below, but the steps are to: 
\begin{enumerate}
\setlength{\itemsep}{0pt}
\item remove the long-timescale variability;
\item measure the initial periodic signal;
\item select the light curve with the optimal photometric aperture;
\item remove periodic stellar signals from the optimal-aperture light curve;
\item infer and correct for instrumental signatures;
\item refine the initial period measurement and consider secondary periodic signatures. 
\end{enumerate}

This procedure, which is similar to that used by \citet{kraus2015} to remove instrumental and out-of-eclipse variability from the {\it K2} light curve for UScoCTIO 5, a spectroscopic binary in the Upper Scorpius star-forming region,
is illustrated in Figure~\ref{fig:lc}. 

We also test a modified version of the \citet{vanderburg2014} algorithm.  Unlike the original \citet{vanderburg2014} algorithm, the modified version removes as much periodic signal as possible before fitting for the instrumental drift. While this method produced light curves consistent with our own, there were some differences. First, our method rejects outliers, and is therefore less sensitive to transits or flares. Second, the modified \citet{vanderburg2014} algorithm introduces jumps in the light curves for stars with longer $P_{rot}$. Our method also more aggressively removes long-term trends from the light curves, but these trends are not periodic and likely not from rotation. 

\subsubsection{Removing long-timescale variability}

We begin by computing a low-power smoothed version of the raw light curve using {\it supersmoother}, a Python implementation of the SuperSmoother non-parametric adaptive smoothing algorithm \citep{friedman1984}.\footnote{\url{https://github.com/jakevdp/supersmoother}} SuperSmoother selects the best local smoothing window for each point in a data series. The degree of smoothing can be set using the bass-enhancement value $\alpha$: low values of $\alpha$ retain more local variability, while high values of $\alpha$ only retain long term trends. We set the bass-enhancement value to the maximum possible value of $\alpha=10$. In most cases, the smoothed curve is effectively linear, but in some cases the data include slow trends, and this method accounts for these trends. This smoothed curve is overlaid on the raw light curve in the top panel of Figure \ref{fig:lc}. 
We divide the data by this smoothed curve to produce the detrended light curve, shown in green in the second panel of Figure \ref{fig:lc}. 

\subsubsection{Measuring the initial periodic signal}

We measure rotation periods using the fast Lomb-Scargle function from the {\it gatspy} package.\footnote{\url{https://github.com/astroML/gatspy}} This is an implementation of the FFT-based algorithm from \citet[][]{press1989}.
We compute the periodogram power for 3$\times$10$^4$ periods ranging from 0.1~d to 70.8~d (the length of the Campaign), though in practice we only consider periods $<$40~d in our analysis. 

At every step, we calculate minimum significance levels for periodogram peaks using non-parametric bootstrap resampling. 
We generate 1000 simulated light curves for each source by holding the observation times fixed and randomly re-drawing the flux values with replacement.
We compute a periodogram for each simulated light curve, and record the maximum power in each periodogram.
The 99.9-percentile value of the maximum peak serves as the 0.1\% significance level for peak detection.

\begin{figure*}[p]
\centerline{\includegraphics[width=2\columnwidth]{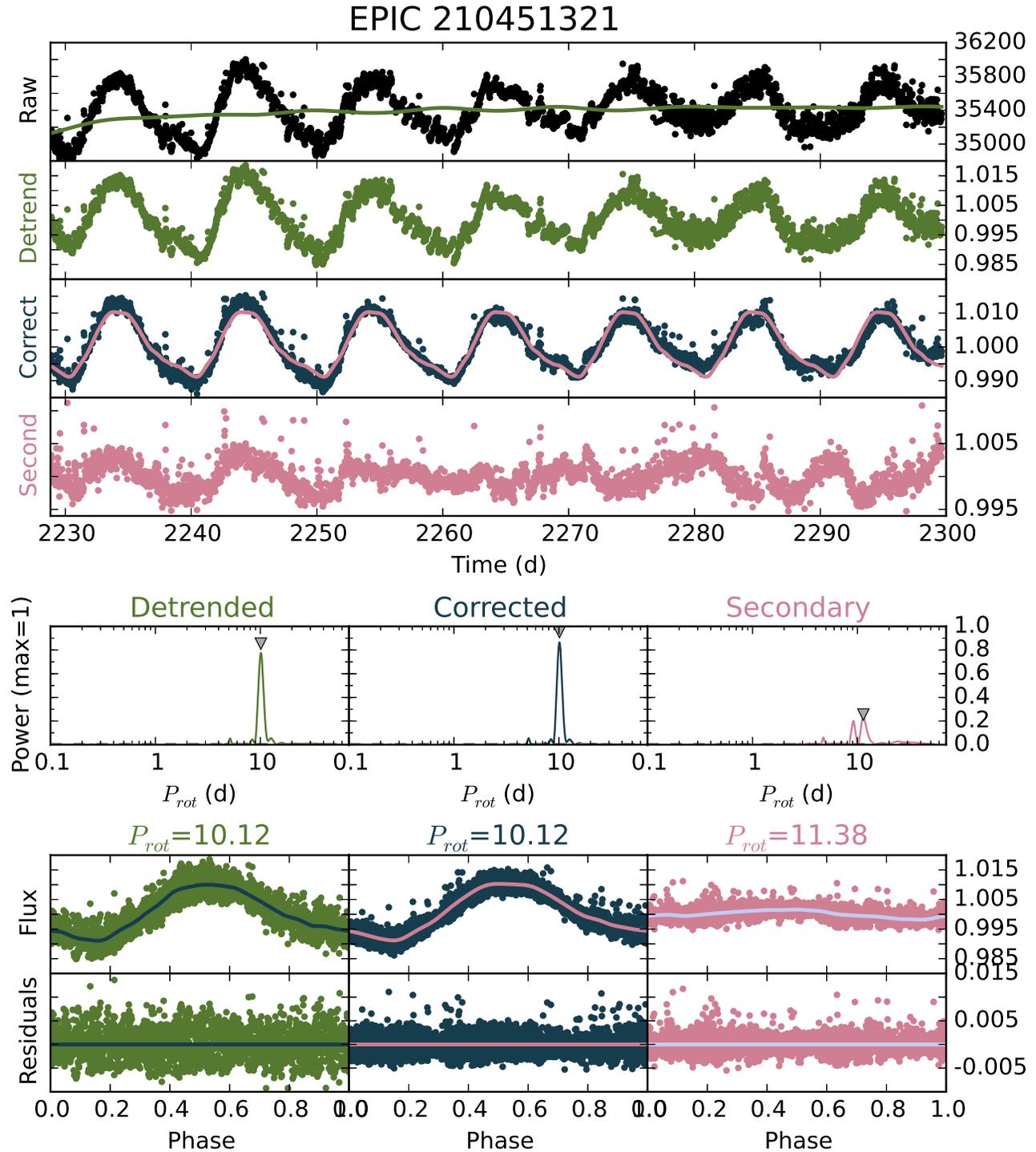}}
\caption{
Analysis of the {\it K2} light curve for EPIC 210451321; plots for all other {\it K2} Hyades targets are found in the electronic edition.
The top four rows show the light curve at different stages of processing. 
The center row shows periodograms for each light curve; the corresponding 99.9\% significance levels (dot-dashed grey line) are generally too close to 0 to be visible. 
The second-to-last row shows the phased light curves with a smoothed version overlaid, and bottom row shows the residuals between those two. 
We divide the raw light curve (top, black) by a smoothed, low-power version (green line) to produce the detrended light curve (second panel, green). 
Either the raw or detrended light curve is chosen for further analysis based on which has the highest periodogram peak between $0.1-40$~d; the corresponding periodogram is shown in the left column. 
We then account for the effect of the spacecraft drift, producing the corrected light curve (blue, third row and center column). 
We also look for additional periods by smoothing the phased, corrected light curve and dividing the periodic signal away, producing the secondary light curve (pink, fourth row and right column). 
In general, the period derived from the corrected light curve is the final period, but in a few cases, the period from the secondary light curve is more likely the true period. 
}
\label{fig:lc} 
\end{figure*}

In general, the minimum significance levels for our periodogram are extremely low. The grey line representing the 0.1\% significance level is rarely visible in the periodograms in Figure~\ref{fig:lc}. 
This is likely due to {\it K2}'s high observation cadence. 
If a periodic signal is present, randomly reshuffling the data will almost certainly destroy it, leaving no significant peaks in the periodogram (the 30-min observing cadence is outside our search range).
Even the 6-hour thruster firing timescale will be wiped out in this resampling. 
Therefore, most periodic signals we find in our analysis are likely real, though we must still distinguish between stellar and instrumental sources. 

We compute periodograms for both the raw and the detrended light curves, and select for further analysis the light curve with the highest peak in its periodogram. 
For this comparison, we only consider peaks at less than 40~d that do not correspond to a low multiple of the 6-hour thruster firing period; the periodogram peaks that fail these cuts likely correspond to the instrumental signatures or long trends in the raw light curve that we are trying to remove. 

\subsubsection{Selecting the optimal photometric aperture}

Initially, we detrend all raw light curves for each target using the process described above, and then select from among all possible apertures the light curve with the highest periodogram peak after the raw/detrended selection. An aperture that is too small will not contain a star's full PSF, decreasing the signal to noise, while a larger aperture may include nearby stars that contaminate the light curve. By selecting the light curve with the strongest initial peak, we begin with the cleanest possible light curve. We refer to the light curve selected for further analysis as the initial light curve.

In some cases, we restrict the maximum possible aperture to exclude nearby stars; see Section \ref{neighbors} for details.

\subsubsection{Removing periodic stellar signals}

We iteratively smooth the initial light curve selected in the previous step before correcting for the centroid drift across the detector. 
We phase-fold the light curve on the best period, and use {\it supersmoother} again to produce a smoothed version of the periodic signal. We then divide the light curve by the smoothed periodic light curve. 
We repeat this process up to six times, or until one of two conditions is met: either the maximum periodogram power falls below 10\% of the initial power, or the highest remaining peak is a harmonic of the 6-hour thruster firing period. This produces a flattened light curve that should only contain instrumental signals. 

\subsubsection{Inferring and correcting for instrumental signatures}

Once all significant periodic signals have been removed from the light curve, leaving it relatively flat, we use it to correct the initial light curve for instrumental signals. At each epoch in the light curve, we find the 21 other epochs with the closest centroid positions (in detector coordinates) to the epoch being corrected. 
We then divide each point in the initial light curve by the median of the flattened light curve at those 21 neighboring points.  This produces the corrected light curve shown in dark blue in the third panel of Figure \ref{fig:lc}; for most of our targets, the $P_{rot}$ measured from the corrected light curve is selected as the final $P_{rot}$.

\begin{figure}[t!]
\centerline{\includegraphics[width=\columnwidth]{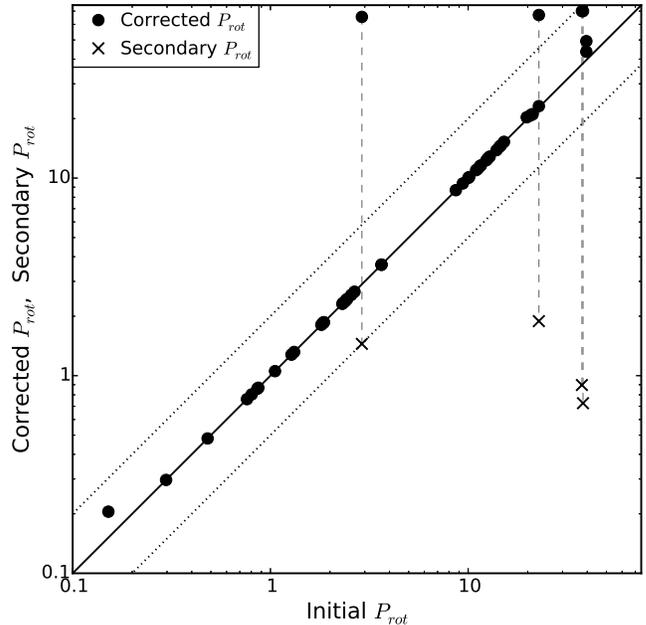}}
\caption{Comparison of $P_{rot}$ measured at different stages of our analysis. The solid line corresponds to a one-to-one match; the  dotted lines represent double- and half-period harmonics. In general, the $P_{rot}$ measured from the initial and corrected light curves are consistent (black dots). In four cases, the corrected $P_{prot}$ is extremely long, and likely represents long-term but non-periodic noise in the corrected light curve. Lines connect each of these $P_{rot}$ values to the crosses that represent the secondary $P_{rot}$, which is more likely the true period (see also Figure \ref{fig:sec}). 
}
\label{fig:comp} 
\end{figure}

The $P_{rot}$ measured from the initial and corrected light curves are in general consistent with each other (Figure \ref{fig:comp}). There are four exceptions, where the $P_{rot}$ measured from both the initial and corrected light curves are spurious, representing non-periodic trends. 

\begin{figure*}[t]
\centerline{\includegraphics[width=2\columnwidth]{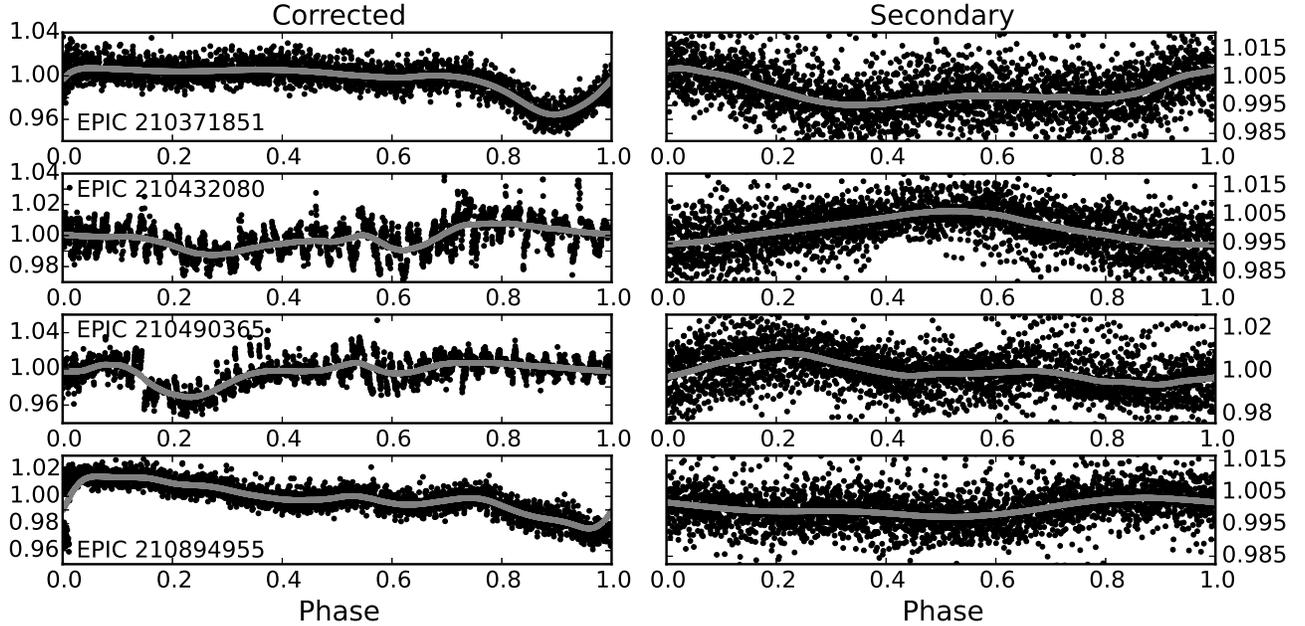}}
\caption{Four light curves for which the $P_{rot}$ measured from the secondary light curve is more likely the true value. The corrected light curves (left) show non-periodic structures that dominate the periodogram. This is removed by dividing out the smoothed curve (grey), which produces the secondary light curves (right).}
\label{fig:sec} 
\end{figure*}

\subsubsection{Refining the initial period measurement}

After correcting for instrumental effects, we also remove the dominant periodic signal to search for any additional signals. We phase-fold the corrected light curve and then use {\it supersmoother} to produce a smoothed version of the periodic signal, shown in pink in the third panel of Figure \ref{fig:lc}. We then divide the corrected light curve by this smoothed periodic curve to produce the secondary light curve, shown in the fourth panel of Figure \ref{fig:lc}. 

We examine by eye the light curves produced for each object at each stage in the analysis. We note quality flags, any evidence of spot evolution and/or multiple periods, and select the final $P_{rot}$ from either the corrected or secondary light curve.

Our visual inspection of the periodograms and phased light curves prevents instrumental signals from contaminating our sample. 
We tested our algorithm on 3600 other Campaign 4 targets with $9\le K_p\le16.5$, and we detect periods between 20-25~d in $\approx20$\% of the corrected light curves. 
These periodogram peaks are typically wide, and appear with other wide peaks around 40~d and 70~d. We believe these to be low frequency noise and/or a harmonic with the 70~d campaign length, and they appear particularly (though not exclusively) in light curves where the spacecraft drift is the only signal. Our visual inspection allows us to separate these spurious detections from genuine sinusoidal rotational variability, and we are confident in the $\approx20$~d rotation periods that we identify for four Hyads.\footnote{See Figures 8.28, 8.32, 8.49, and 8.56 for the light curves of EPIC 210704853, EPIC 210736105, EPIC 210941195, and EPIC 211036776.}

We flag the light-curve quality on a scale of 0--2: 0 is a confident detection, 1 indicates significant residual noise in the light curve that may affect the final period measurement, or confusion about which periodogram peak corresponds to the dominant period, and 2 indicates that the spacecraft drift pattern dominates the light curve or that there is some other fatal problem with the period measurement. 

We also add flags for the presence of spot evolution and multiple periodic signals in the light curve. For spot evolution and multiperiodic signals, we use ``Y,'' ``M,'' and ``N'' flags, for ``Yes,'' ``Maybe,'' and ``No.'' When there is significant power at half or twice the chosen period, we phase-fold the light curve at those harmonics and note in Table \ref{tab:k2} whether they could be the true period. We also note in Table \ref{tab:k2} when there are two clear peaks in the periodogram and there is confusion over which is the dominant period, or when any visible periodic signals are not picked up in our analysis.

In most cases, the secondary period is a harmonic of either the thruster-firing period or the fundamental period, or a clear, unrelated signal. In four cases, however, the period from the corrected light curve is spurious, and the period from the secondary light curve is selected as the real period. These four corrected light curves, shown in Fig. \ref{fig:sec} show structure that was not removed by the initial low-power detrending, but which was removed by the smoothing used to produce the secondary light curve. The periods measured from the initial and corrected light curves appear to be an instrumental signal, and the secondary period looks more like a rotational signal (see Figures \ref{fig:comp} and \ref{fig:sec}). 

\subsection{Light Curves With Multiple Periodic Signals}

A subset of Hyades light curves in our sample show multiple significant periodogram peaks. The secondary peaks can generally be classified into two categories: harmonics of the fundamental period that appear due to symmetric spot coverage or spot evolution (21 cases) or a real secondary period (eight cases).

Spot evolution and double-dip light curves due to symmetric spot distributions have been observed in many {\it Kepler} stars \citep[e.g.,][]{walkowicz2013, mcquillan2013}. All 11 Hyads with $P_{rot}$ from the literature and this work show at least some evidence for spot evolution in their {\it K2} light curves. In two of these cases we find a different rotation period than in the literature (see Section \ref{consistency} below), and spot evolution is the likely culprit for this discrepancy. 

The cause of two distinct signals in the light curve is more difficult to untangle, but likely comes from an unresolved second star on the chip. Of the 11 stars with light curves that definitely or possibly contain multiple periodic signals, seven are confirmed binaries and one has a neighbor that is unresolved on the {\it K2} chip.

\begin{deluxetable*}{ccccccl}[t]
\tablewidth{0pt}
\tabletypesize{\scriptsize}
\tablecaption{{\it K2} targets for which $P_{rot}$ could not be measured \label{tab:k2bad}}
\tablehead{
\colhead{}& \colhead{}& \colhead{}& \colhead{{\it Kp}}& \colhead{Prev.}& \colhead{}& \colhead{}\\ 
 \colhead{EPIC}& \colhead{RA}& \colhead{Dec}& \colhead{(mag)}& \colhead{$P_{rot}$ (d)\tablenotemark{a}}& \colhead{Bin\tablenotemark{b}}& \colhead{Notes}
}
\startdata
210643507 & 03:40:45.81 & 17:34:47.77 & 16.64 & -& 0& 
noise \\
210563410 & 03:42:04.44 & 16:25:18.29 & 15.13 & -& 0& 
periodicity in large aperture light curve, may be from neighbor\\
211084463 & 03:49:42.70 & 24:19:07.80 & 14.99 & -& 0& 
much brighter blended neighbor\\
210359769 & 03:55:01.44 & 12:29:08.10 & 9.76 & D& 0& 
periodicity in large aperture light curve, may be from neighbor\\
210384590 & 03:58:35.82 & 13:06:18.11 & 9.17 & -& 1& 
noise; unclear if true stellar signal\\
210976212 & 03:59:51.83 & 22:36:13.45 & 14.41 & -& 0& 
noise\\
210835395 & 04:04:27.04 & 20:24:30.43 & 15.16 & -& 2& 
period close to firing period\\
210827030 & 04:07:03.25 & 20:16:50.87 & 13.81 & -& 0& 
period close to firing period\\
211037886 & 04:14:32.32 & 23:34:29.80 & 9.05 & A& 0& 
saturated and bleeds out of pixel stamp (estimated $K_p$ is too faint)\\
211088189 & 04:14:50.19 & 24:22:37.52 & 16.02 & -& 0& 
noise\\
210517837 & 04:15:33.69 & 15:42:22.57 & 10.46 & -& 2& 
noise\\
210707811 & 04:17:50.63 & 18:28:30.69 & 12.81 & -& 2& 
periodicity in large aperture light curve, may be from neighbor\\
210693497 & 04:22:39.56 & 18:16:09.61 & 12.13 & -& 0& 
periodicity in large aperture light curve, may be from neighbor\\
210754620 & 04:24:20.95 & 19:10:50.66 & 14.40 & -& 0& 
noise\\
210742017 & 04:24:30.59 & 18:59:12.82 & 15.15 & -& 2& 
periodicity in large aperture light curve, may be from neighbor\\
210667985 & 04:25:25.03 & 17:54:55.23 & 10.62 & -& 2& 
saturation strip from brighter star crosses PSF\\
210743724 & 04:25:41.84 & 19:00:47.57 & 13.80 & -& 2& 
periodicity in large aperture light curve, may be from neighbor\enddata
\tablecomments{ For details on the light curve analysis of these stars, see the electronic version of Table \ref{tab:k2}.}
\tablenotetext{a}{Source of previous $P_{rot}$ measurement: ``D'' for \citet{delorme2011} and ``A'' for our ASAS analysis (Cargile et al. in prep).}
\tablenotetext{b}{Binary status (Section \ref{binaries}, Table \ref{tab:bin}): 0 indicates a single star (to the best of our knowledge), 1 is a photometrically identified candidate binary, and 2 is a confirmed binary from the literature.}
\end{deluxetable*}

\section{Results}\label{res}

\begin{figure}[t!]
\centerline{\includegraphics[width=\columnwidth]{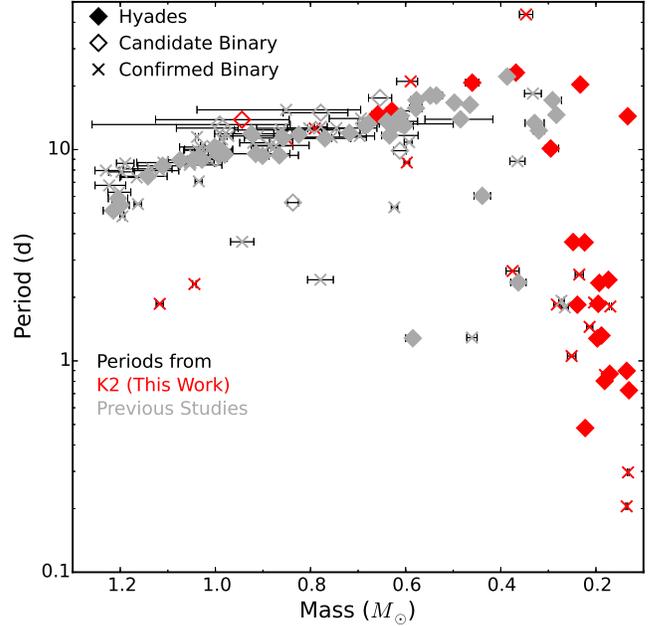}}
\caption{Mass-period distribution for all Hyads with measured periods. Grey symbols represent literature $P_{rot}$, and red symbols denote new $P_{rot}$ measured from {\it K2} data. Empty diamonds and crosses denote photometrically identified candidate binaries and confirmed binaries, respectively. 
}
\label{fig:periodmass_k2} 
\end{figure}

We present $P_{rot}$ measurements for 48 Hyades members, including 37 members with no prior $P_{rot}$ measurement. Our $P_{rot}$ values, flags, and analysis outputs are found in Table~\ref{tab:k2}, where we also note whether we initially choose the raw or detrended light curve for analysis, and whether the final period is selected from the corrected or secondary light curve.  Our new rotation periods, along with literature values, are shown as a function of stellar mass in Figure~\ref{fig:periodmass_k2}.

Nearly all of these new periods are for Hyads with $M_*\ \lapprox\ 0.7\ \Msun$, and the majority are for fully convective, $\lapprox$0.3~\Msun\ Hyads. These are the first $P_{rot}$ measurements for a set of Hyades members with $0.1\ \lapprox\ M_*\ \lapprox\ 0.3\ \Msun$. 

We could not measure $P_{rot}$ for the 17 {\it K2} targets listed in Table~\ref{tab:k2bad}. In two cases, nearby bright stars contaminate the PSF. The PSF of another target extends beyond the pixel stamp; the estimated $K_p$ magnitude was likely too faint. 
Six other stars with nearby neighbors show periodic signals in light curves extracted with apertures of $r=4-6$~pixels, but exhibit mostly noise when the aperture is restricted to exclude these faint neighbors. Thus, it is possible that a neighbor is the source of the periodic variability rather than the target star. 
Finally, in two cases, the signal looks sinusoidal, but the measured $P_{rot}$ is too close to the 6-hour thruster firing period to be sure that it is a stellar signal. 

For the remaining six stars, however, the light curves are dominated by instrumental noise caused by the pointing drift. Although some of our faintest targets are among these 16, the noise-dominated light curves come from stars across the full brightness range of our sample. Thus, these noisy stars are likely to have very small amplitude variability, and our inability to extract periods for these stars does not necessarily imply a failure of our method for faint stars.  

\subsection{Consistency With Prior $P_{rot}$ Measurements}\label{consistency}

Eleven {\it K2} targets have $P_{rot}$ measured by previous authors; the measurements are generally consistent (see Table~\ref{tab:protdiff}), with two exceptions.

\citet{delorme2011} measure $P_{rot}=13.51$~d for EPIC 210771915, while our analysis of ASAS data gave us 
$P_{rot}=7.19$~d. In D14, we used the shorter ASAS-derived period, but 
our analysis of the {\it K2} data shows that the longer \citet{delorme2011} period is correct. The {\it K2} light curve for this star shows a prominent double-dip feature and evidence for spot evolution. The shorter period we initially derived 
is a half-period harmonic of the true period. 

In our analysis of ASAS stars, we measure $P_{rot}=8.37$~d for EPIC 211058178, while we measure $P_{rot}=12.32$~d with {\it K2} data. The {\it K2} light curve for this star shows rapid evolution on timescales close to and shorter than the rotation period, and the amplitude of variability drops from nearly 1\% to $\approx$0.1\% over the course of Campaign 4. The star is also a confirmed spectroscopic binary \citep{paulson2004}, and is therefore excluded from our gyrochronology analysis. 

Overall, however, the $P_{rot}$ measurements are consistent from study to study. For example,
one \citet{prosser1995} target was also observed by \citet{delorme2011} and by {\it K2}, and the three derived $P_{rot}$ agree at the 1\% level. The variation among $P_{rot}$ measured by different surveys for the same star are frequently of order a few percent, and usually $<$10\% (see Table~\ref{tab:protdiff}). This overall consistency is a good sign for future rotation studies, especially since {\it K2}'s baseline and cadence do allow us to identify two instances where sampling likely affected earlier results. 


\subsection{Stellar Rotation and Multiplicity at 625~Myr}


Based on previous studies, we expect Hyades members with $M_*\ \gapprox\ 0.6-0.7\ \Msun$ to be confined to a single-valued $M_*(P_{rot})$ relation that also defines the upper envelope of rotation for lower-mass stars, while less massive stars to show a wide spread in $P_{rot}$ \citep{barnes2003, irwin2009, agueros11}. We refer to the upper envelope of $P_{rot}$ values as the slow-rotator sequence, and the lower envelope as the rapid rotator sequence \citep[labelled  the ``I sequence'' and ``C sequence,'' respectively, by][]{barnes2003}. We also expect all fully convective stars to be rapidly rotating at this age, with a break in the slow-rotator sequence around the transition between partially and fully convective stars \citep{irwin2009, newton2015}. 

The upper envelope of rotation in the Hyades is consistent with previous observations of rotation in open clusters. One notable exception, EPIC 210489654 ($M_*=0.35$~\Msun, $P_{rot}=45$~d), lies well above the rest of the Hyads. It is a confirmed binary \citep{janson2014}, which may be the cause of its unusually slow $P_{rot}$ for this age, or it may be an older field star whose proper motion happens to match the Hyades's. Besides this outlier, the Hyades slow-rotator sequence is well-defined for $M_*\ \gapprox\ 0.3\ \Msun$ stars, with $P_{rot}$ increasing slightly toward lower masses. 

Once binary systems are excluded, we do not see strong evidence for a transition in rotation properties at $M_*\approx0.6\ \Msun$. 
Only three rapid rotators with $0.6\gtrsim M_*\gtrsim0.3\ \Msun$ remain, and none has been surveyed for binary companions. Therefore, we cannot rule out multiplicity \citep{morgan2012} or closely orbiting giant planets \citep{katja2014} as the cause for their more rapid rotation. These outliers should be priority targets for future multiplicity surveys; they are most likely binaries, meaning that the period-mass relation for single stars is single-valued for $M_* \gtrsim 0.3\ \Msun$

It is also possible that the reported $P_{rot}$ for these three remaining rapid rotators with $0.6\gtrsim M_* \gtrsim0.3\ \Msun$ are aliases of longer periods. All three have $P_{rot}$ derived from the same survey \citep{delorme2011}, and these authors note that they are sensitive to $P_{rot}$ between $1-20$~d. 
Although only a few {\it K2} targets fall within the $0.6 \gtrsim M_* \gtrsim 0.3\ \Msun$ range, we are sensitive to the full range of possible $P_{rot}$ values. 
However, we only detect long ($\gapprox10$~d) rotation periods for stars in this mass range. 

The break in the slow-rotator sequence around the transition to full convection ($\approx$0.3~$\Msun$) is hard to quantify due to a pair of slowly rotating mid-M dwarfs. EPIC 211137806 ($M_*=0.13$~\Msun, $P_{rot}=14.38$~d) is newly identified as a candidate Hyad by \citet{roser2011}. 
EPIC 210736105 ($M_*=0.23$~\Msun, $P_{rot}=20.31$~d) was cataloged as a cluster member by \cite{reid1992}, but it has not been surveyed for multiplicity.\footnote{EPIC 210736105 also has an angular light curve with a fast rise and slow decline reminiscent of a Cepheid variable (Figure~\ref{fig:neighbor}). 
However, period-luminosity relationships for Classical or Type II Cepheids imply a distance of $\approx$24~kpc or $\approx$7~kpc, respectively \citep{gieren1998, matsunaga2009}, which is too far away to account for the observed proper motion. There is also a brighter star nearby, but that star does not show evidence of periodic variability (Figure~\ref{fig:neighbor}).} 
Based on their slow rotation periods, these stars are likely either unresolved multiple systems or older field interlopers, but we cannot confirm this without radial velocity or high-resolution imaging data. 

Despite these two outliers, nearly all Hyads with $M_*\ \lapprox\ 0.3~$\Msun\ have $P_{rot}<5$~d. This is consistent with existing evidence that fully convective M dwarfs stay spinning rapidly for several Gyr \citep{newton2015}. The fastest fully convective rotators are confirmed binary systems, similar to the effect observed in more massive stars. 

\section{Discussion}\label{disc}

\subsection{Implications for Gyrochronology
}\label{gyro}

As described above, empirical studies of rotation in open clusters 
identify two regimes in the $P_{rot}(M_*)$ plane for partially convective stars: 
above a particular mass, most stars are confined to a single-valued $P_{rot}(M_*)$ sequence, and below this mass, stars exhibit a wide range of $P_{rot}$ values \citep{barnes2003}. Gyrochronology studies typically focus on the evolution of the slow-rotator sequence, which is unaffected by our results.  The same studies make predictions for the age-dependent location of the transition between the single-valued sequence and the wider spread in $P_{rot}$ values.  Previous work on the Hyades and the co-eval Praesepe cluster suggests that this transition occurs around $0.6-0.7$ \Msun\ for $\approx$600~Myr-old stars \citep[][D14]{agueros11}.

Our results indicate that, after confirmed binaries are removed, the sequence of slowly rotating likely single-star Hyades members breaks down at $\approx$0.3~\Msun.
This is has important consequences for models of stellar spin-down.
If single $\gapprox$0.3~\Msun\ stars converge onto the slow-rotator sequence by the age of the Hyades, then this will skew empirical gyrochronological relations \citep[i.e.,][]{barnes2003}, which are currently calibrated such that this transition occurs at $\approx$0.6~\Msun\ in the Hyades. 

Since rotation and activity are generally correlated, we would expect that any behavior observed in the mass-period plane would be reflected when analyzing the strength of the \halpha line (a chromospheric proxy for the magnetic-field strength) for these same stars. Specifically, if rapid rotators $\gapprox$0.3 \Msun\  are all binaries, then \halpha emitters above that threshold mass should predominantly be binaries as well. 
In D14 we found that, for stars in the Hyades and the co-eval Praesepe cluster, both $P_{rot}$ and \halpha activity transitioned from a mostly single-valued sequence to a wider spread in $P_{rot}$ or equivalent width at about $0.6-0.7$ \Msun, or spectral types K7-M0. However, we did not consider the binary status of Praesepe and Hyades members in that analysis. 

Because binary companions will add excess $K$-band flux and impact our mass calculations, we now re-examine this transition as a function of $(r'-K)$  rather than mass. Using color keeps equal-mass binaries closer to the correct location for their spectral type, although low-mass companions could still redden a star's colors. 

\begin{figure}[!t]
\centerline{\includegraphics[width=\columnwidth]{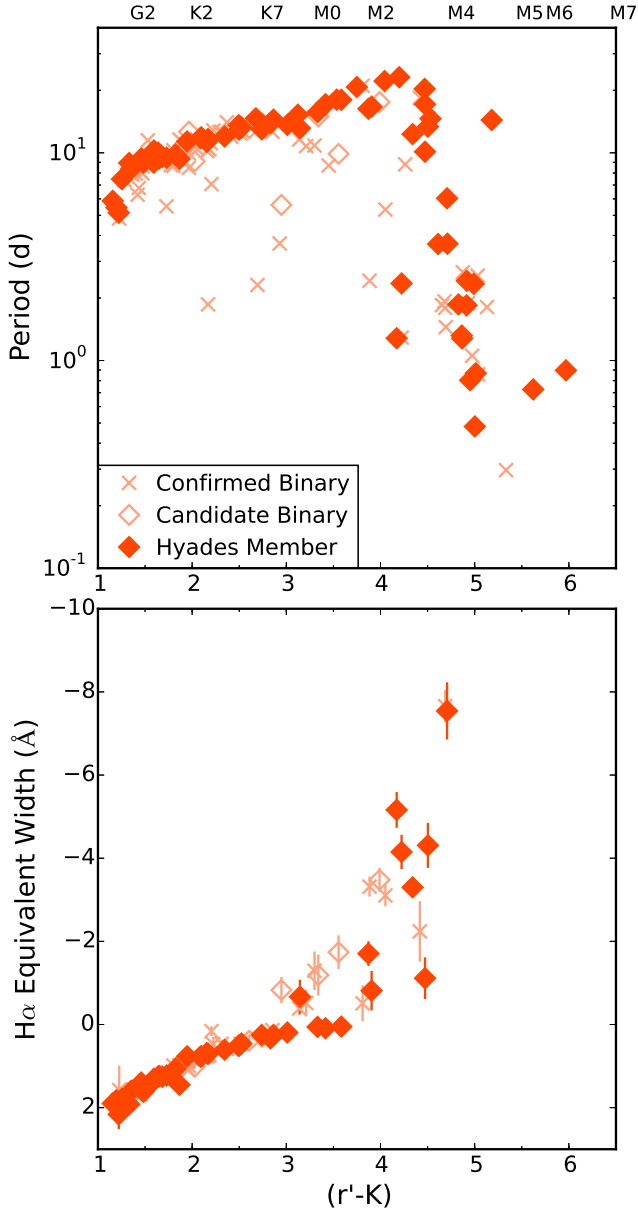}}
\caption{{\it Top} --- $P_{rot}$ as a function of color for all Hyads with measured periods. Empty diamonds and crosses denote photometrically identified candidate binaries and confirmed binaries, respectively.  {\it Bottom} --- \halpha emission as a function of color for Hyads with optical spectra in D14. Negative equivalent widths indicate emission. We find the same pattern in activity that we find in rotation. Most stars bluer than $(r'-K)\approx4$ lie along a sequence of slow rotation/low activity, with only a few rapid rotators or active stars among them. Most of these outliers are confirmed or candidate binaries. Redward of $(r'-K)\approx4$, both $P_{rot}$ and equivalent width values show much more scatter.
}
\label{fig:activity} 
\end{figure}

We find that the transition in activity and $P_{rot}$ occurs around $(r'-K)\approx4$, corresponding to spectral types of M2-M3 or $M_*\approx0.3-0.4\ \Msun$ (Figure \ref{fig:activity}). Bluer stars with higher-than-average \halpha equivalent widths or rotating more rapidly than average are nearly all binaries. 
This provides further evidence that binaries have contaminated previous studies of $P_{rot}$ evolution in open clusters. 

The Hyades's proximity means it has been extensively studied for multiplicity, and may be the best current example of how binaries affect the period-mass plane. Other, more distant clusters may not have received the same attention.  Roughly 45\% of Hyads with measured $P_{rot}$ are confirmed binaries, and in total $\gapprox$30\% of all Hyads are confirmed binaries.  The co-eval Praesepe cluster, in contrast, has $\approx$1200 members \citep{adam2007} compared to the Hyades's $\approx$800, but <20\% Praesepe stars have confirmed companions. While experiments like {\it K2} are providing many new $P_{rot}$ measurements for open cluster stars, complementary binary surveys are needed to investigate the impact of multiplicity on current age-rotation relations.

\subsection{Implications for Magnetic Braking Models}\label{brake}

\begin{figure}[t]
\centerline{\includegraphics[width=\columnwidth]{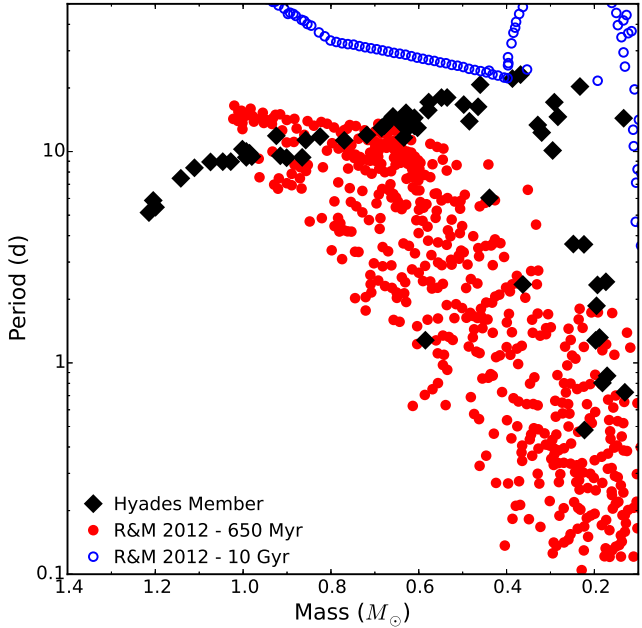}}
\caption{Mass-period distribution for single Hyads (black diamonds; includes both confirmed single stars and stars that have not been surveyed for multiplicity) compared to \citet{reiners2012} model data for 650-Myr-old (red closed circles) and 10-Gyr-old (blue open circles) stars.}
\label{fig:reiners} 
\end{figure}

\citet{reiners2012} derive a model for the angular momentum lost by a rotating solid sphere due to magnetic braking. These authors initialize their model using $P_{rot}$ and masses measured for 5-Myr-old stars in the Orion Nebula Cluster and NGC 2264, then evolve these stars forward in time. They show that their model broadly reproduces the shape of the mass-period plane at 625~Myr as defined by then-available $P_{rot}$ for Praesepe and the Hyades, but a detailed comparison to these data shows several discrepancies. The upper envelope of slow rotators in the \citet{reiners2012} model has a different slope than the data; these authors attribute this to the effects of core-envelope coupling, which their model does not include. The \citet{reiners2012} model also predicts shorter $P_{rot}$ values than is observed for $M_*\ \lapprox\ 0.6\ \Msun$ (Figure~\ref{fig:reiners}).

\begin{figure}[t]
\centerline{\includegraphics[width=\columnwidth]{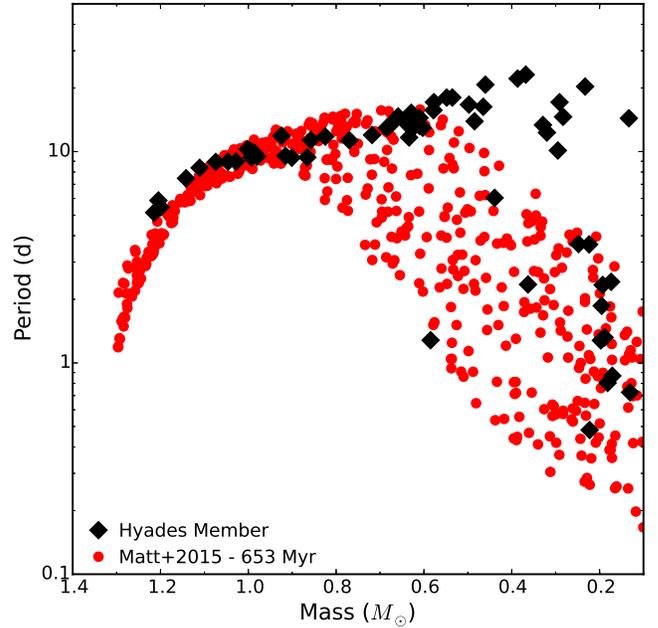}}
\caption{As in Figure~\ref{fig:reiners}, but the red circles are \citet{matt2015} model data for 653-Myr-old stars.}
\label{fig:matt} 
\end{figure}

\citet{matt2015} also derive a model for the spin-down of a solid sphere due to magnetic braking. These authors' initial conditions approximate the distribution of $P_{rot}$ observed for $2-5$~Myr-old stars, but are not drawn directly from observations. They also allow the stellar radius and moment of inertia to evolve with time following evolutionary tracks. The \citet{matt2015} model better predicts the mass dependence of the slow-rotator sequence for Hyads with $M_*\ \gapprox\ 0.6\ \Msun$ (Figure~\ref{fig:matt}). 


\citet{matt2015} find that their model at 573~Myr fails to predict a set of rapid rotators in Praesepe with $0.7\ \lapprox\ M_*\ \lapprox\ 1.0\ \Msun$, and these authors attribute the discrepancy to a change in the magnetic torque exerted on these stars. However, we find that any rapidly rotating Hyads in that mass regime are known multiple systems, and most of the rapidly rotating Praesepe members at those masses are identified as candidate binaries in D14.

Both models predict more rapid rotators than observed for $\lapprox$0.9~\Msun\ Hyads, which indicates that magnetized winds more effectively brake stellar rotation than the models predict. These discrepancies are most pronounced for $M_*\ \lapprox\ 0.6\ \Msun$, and suggest two transition points in braking behavior: one around the beginning of the M-dwarf regime, and one at the boundary to full convection.

Hyads with $0.3\ \lapprox\ M_*\ \lapprox\ 0.6\ \Msun$ rotate more slowly than either model predicts, extending the slow-rotator sequence to the fully convective boundary.  \citet{matt2015} suggest that this is due to internal angular momentum transport that is not included in their models.

Alternately, a change in magnetic field geometry could increase the braking efficiency for M dwarfs relative to the prescriptions of \citet{matt2015}, which match G and K stars well. Zeeman Doppler imaging studies find a transition in magnetic field properties at around 0.5~\Msun, not at the transition to fully convective stars \citep{donati2009, donati2011}. Most M dwarfs have much stronger and more poloidal fields than their more massive counterparts with the same $P_{rot}$, though some still have weaker and/or more toroidal fields. Poloidal fields should lead to more effective magnetic braking because they allow for larger magnetic wind torques due to a larger Alfv\'en radius and more efficient mass loss in the equatorial plane, compared to more complex fields \citep{garraffo2015, garraffo2015-1}. This could explain why the models fail to reproduce the population of slow rotators with $0.3\ \lapprox\ M_*\ \lapprox\ 0.6\ \Msun$ in the Hyades and Praesepe; future models will need to account for the enhanced braking efficiency in this mass range.  

A second transition point occurs around the fully convective boundary, 
where core-envelope coupling becomes irrelevant. Since these stars have no core to decouple from their convective zone, the solid-body model used by \citet{reiners2012} and \citet{matt2015} should match these $\lapprox$0.3~\Msun\ stars best, and the stellar wind model is a likely culprit for the observed discrepancy.
The discrepancy with observations is more pronounced in the \citet{reiners2012} model, where the model $P_{rot}$ distribution appears to be shifted down by a factor of $3-4$ relative to observations. The \citet{matt2015} model extends to slightly faster $P_{rot}$ than we observe in this mass range, but overall does a better job of replicating the fully convective Hyads.
This indicates that, despite the discrepancy with early M dwarfs, the \citet{matt2015} model better captures the overall internal angular momentum and wind loss rates for fully convective stars.

\section{Conclusions}\label{concl}

\begin{enumerate}
\item We have extracted and analyzed {\it K2} light curves for 65 Hyades members. We measure rotation periods for 48 of these, including 37 Hyads without previous $P_{rot}$. This is the first time rotation periods have been measured for single Hyades members that definitively fall below the fully convective boundary ($M_*\ \lapprox\ 0.3\ \Msun$).  A total of 139 Hyades members now have measured $P_{rot}$. 
\item We have assembled binary information from the literature for all Hyads with measured $P_{rot}$ and for those targeted by {\it K2}. We find that 63 of 139 stars with measured $P_{rot}$ are confirmed binary or multiple systems. This ensures that, to the best of present knowledge, our gyrochronological analysis is based on a sample of single stars.
\item We find that most, if not all, rapid rotators with $M_*\ \gapprox\ 0.3\ \Msun$ are multiple systems. Most single stars have spun down to a single-valued mass-period relationship, indicating that multiplicity must be considered when analyzing the mass-period plane.
\item We find that magnetic braking models predict more rapid rotation at this age than observed over a range of masses. The discrepancy in the fully convective regime in particular indicates that magnetic braking is more efficient than predicted. 
\end{enumerate}

The Hyades's large apparent size on the sky has limited surveys for rotation in the cluster, and the mass range where rotators appear to be missing is at the faint end of previous surveys. This is compounded by a dip in membership numbers around $0.4\lesssim M_* \lesssim 0.6\ \Msun$. Even if our data suggest more fundamental theoretical challenges, at this time we cannot rule out these selection effects as being responsible for points 3--4 above.

Additional $P_{rot}$ measurements, both for the Hyades and the co-eval Praesepe cluster, are needed to rule out these selection effects. {\it K2} observed $\approx$700 Praesepe members in Campaign 5; those data are now public, and will be the focus of our next paper. Furthermore, {\it K2} is slated to return to the Hyades in Campaign 13, and the planned field position could yield $\approx$100 additional $P_{rot}$ for cluster members. 

More binary surveys are also needed in both clusters to match these $P_{rot}$ samples and confirm the effect of multiplicity on the period-mass plane at $\approx$600~Myr. The remaining rapidly rotating $M_*\ \gapprox$0.3~\Msun\ Hyads have not been surveyed for companions, and many planned Campaign 13 targets are also missing from existing multiplicity studies. Comparable $P_{rot}$ and multiplicity samples are also needed in other clusters to further investigate how binarity impacts age-rotation relations.


\acknowledgments

S.T.D.~gratefully acknowledges the hospitality of the {\it Kepler}/{\it K2} Science Team at NASA Ames Research Center, particularly the advice and support of Geert Barentsen, Knicole Colon, Mike Haas, Mark Messersmith, Fergal Mullally, and Susan~E.~Thompson.
We thank Sean Matt, Subu Mohanty, and Ansgar Reiners for discussing and sharing their models with us, and Ruth Angus and Dan Foreman-Mackey for useful discussions and advice. 
We also thank Alisha Kundert for her work on the Hyades catalog and ASAS analysis. 
M.A.A.~acknowledges support provided by the NSF through grants AST-1255419 and AST-1517367. K.R.C.~acknowledges support provided by the NSF through grant AST-1449476. 
We thank the anonymous referee for their critique, which improved the paper.

Our work has greatly benefited from the {\it supersmoother} and {\it gatspy} packages developed by J.~Vanderplas\footnote{\url{https://github.com/jakevdp/supersmoother}, \\ \url{https://github.com/astroML/gatspy}} and the {\it pywcsgrid2} package developed by J.~Lee.\footnote{\url{https://github.com/leejjoon/pywcsgrid2}} We have also made use of the {\it Astropy} package \citep{astropy}, especially the affiliated {\it photutils} package.\footnote{\url{http://photutils.readthedocs.org/en/latest/}} 

This research has made use of NASA's Astrophysics Data System Bibliographic Services, the SIMBAD database, operated at CDS, Strasbourg, France, and the VizieR database of astronomical catalogs \citep{Ochsenbein2000}. 

This paper includes data collected by the {\it K2} mission. Funding for the {\it K2} mission is provided by the NASA Science Mission directorate. Some of the data presented in this paper were obtained from the Mikulski Archive for Space Telescopes (MAST). STScI is operated by the Association of Universities for Research in Astronomy, Inc., under NASA contract NAS5-26555. Support for MAST for non-{\it HST} data is provided by the NASA Office of Space Science via grant NNX09AF08G and by other grants and contracts.

This research has made use of the NASA/ IPAC Infrared Science Archive, which is operated by the Jet Propulsion Laboratory, California Institute of Technology, under contract with the National Aeronautics and Space Administration. The Two Micron All Sky Survey was a joint project of the University of Massachusetts and IPAC.  

The Digitized Sky Survey was produced at the Space Telescope Science Institute under U.S. Government grant NAG W-2166. The images of these surveys are based on photographic data obtained using the Oschin Schmidt Telescope on Palomar Mountain and the UK Schmidt Telescope. The plates were processed into the present compressed digital form with the permission of these institutions.

Funding for SDSS-III has been provided by the Alfred P. Sloan Foundation, the Participating Institutions, the National Science Foundation, and the U.S. Department of Energy Office of Science. The SDSS-III web site is http://www.sdss3.org/. SDSS-III is managed by the Astrophysical Research Consortium for the Participating Institutions of the SDSS-III Collaboration.

This work was enabled using the Pan-STARRS 1 System operated by the PS1 Science Consortium (PS1SC) and its member institutions. The PS1 Surveys have been made possible through the combinations of the Institute for Astronomy at the University of Hawaii, The Pan-STARRS Project Office, the Max-Planck Society and its participating institutes.

\setlength{\baselineskip}{0.6\baselineskip}
\bibliography{references}

\begin{thebibliography}{}
\expandafter\ifx\csname natexlab\endcsname\relax\def\natexlab#1{#1}\fi

\bibitem[{{Ag{\"u}eros} {et~al.}(2011){Ag{\"u}eros}, {Covey}, {Lemonias},
  {Law}, {Kraus}, {Batalha}, {Bloom}, {Cenko}, {Kasliwal}, {Kulkarni},
  {Nugent}, {Ofek}, {Poznanski}, \& {Quimby}}]{agueros11}
{Ag{\"u}eros}, M.~A., {Covey}, K.~R., {Lemonias}, J.~J., {et~al.} 2011, apj,
  740, 110

\bibitem[{{Aihara} {et~al.}(2011){Aihara}, {Allende Prieto}, {An}, {Anderson},
  {Aubourg}, {Balbinot}, {Beers}, {Berlind}, {Bickerton}, {Bizyaev}, {Blanton},
  {Bochanski}, {Bolton}, {Bovy}, {Brandt}, {Brinkmann}, {Brown}, {Brownstein},
  {Busca}, {Campbell}, {Carr}, {Chen}, {Chiappini}, {Comparat}, {Connolly},
  {Cortes}, {Croft}, {Cuesta}, {da Costa}, {Davenport}, {Dawson}, {Dhital},
  {Ealet}, {Ebelke}, {Edmondson}, {Eisenstein}, {Escoffier}, {Esposito},
  {Evans}, {Fan}, {Femen{\'\i}a Castell{\'a}}, {Font-Ribera}, {Frinchaboy},
  {Ge}, {Gillespie}, {Gilmore}, {Gonz{\'a}lez Hern{\'a}ndez}, {Gott}, {Gould},
  {Grebel}, {Gunn}, {Hamilton}, {Harding}, {Harris}, {Hawley}, {Hearty}, {Ho},
  {Hogg}, {Holtzman}, {Honscheid}, {Inada}, {Ivans}, {Jiang}, {Johnson},
  {Jordan}, {Jordan}, {Kazin}, {Kirkby}, {Klaene}, {Knapp}, {Kneib},
  {Kochanek}, {Koesterke}, {Kollmeier}, {Kron}, {Lampeitl}, {Lang}, {Le Goff},
  {Lee}, {Lin}, {Long}, {Loomis}, {Lucatello}, {Lundgren}, {Lupton}, {Ma},
  {MacDonald}, {Mahadevan}, {Maia}, {Makler}, {Malanushenko}, {Malanushenko},
  {Mandelbaum}, {Maraston}, {Margala}, {Masters}, {McBride}, {McGehee},
  {McGreer}, {M{\'e}nard}, {Miralda-Escud{\'e}}, {Morrison}, {Mullally},
  {Muna}, {Munn}, {Murayama}, {Myers}, {Naugle}, {Neto}, {Nguyen}, {Nichol},
  {O'Connell}, {Ogando}, {Olmstead}, {Oravetz}, {Padmanabhan},
  {Palanque-Delabrouille}, {Pan}, {Pandey}, {P{\^a}ris}, {Percival},
  {Petitjean}, {Pfaffenberger}, {Pforr}, {Phleps}, {Pichon}, {Pieri}, {Prada},
  {Price-Whelan}, {Raddick}, {Ramos}, {Reyl{\'e}}, {Rich}, {Richards}, {Rix},
  {Robin}, {Rocha-Pinto}, {Rockosi}, {Roe}, {Rollinde}, {Ross}, {Ross},
  {Rossetto}, {S{\'a}nchez}, {Sayres}, {Schlegel}, {Schlesinger}, {Schmidt},
  {Schneider}, {Sheldon}, {Shu}, {Simmerer}, {Simmons}, {Sivarani}, {Snedden},
  {Sobeck}, {Steinmetz}, {Strauss}, {Szalay}, {Tanaka}, {Thakar}, {Thomas},
  {Tinker}, {Tofflemire}, {Tojeiro}, {Tremonti}, {Vandenberg}, {Vargas
  Maga{\~n}a}, {Verde}, {Vogt}, {Wake}, {Wang}, {Weaver}, {Weinberg}, {White},
  {White}, {Yanny}, {Yasuda}, {Yeche}, \& {Zehavi}}]{DR8paper}
{Aihara}, H., {Allende Prieto}, C., {An}, D., {et~al.} 2011, The Astrophysical
  Journal Supplement Series, 193, 29

\bibitem[{{Astropy Collaboration} {et~al.}(2013){Astropy Collaboration},
  {Robitaille}, {Tollerud}, {Greenfield}, {Droettboom}, {Bray}, {Aldcroft},
  {Davis}, {Ginsburg}, {Price-Whelan}, {Kerzendorf}, {Conley}, {Crighton},
  {Barbary}, {Muna}, {Ferguson}, {Grollier}, {Parikh}, {Nair}, {Unther},
  {Deil}, {Woillez}, {Conseil}, {Kramer}, {Turner}, {Singer}, {Fox}, {Weaver},
  {Zabalza}, {Edwards}, {Azalee Bostroem}, {Burke}, {Casey}, {Crawford},
  {Dencheva}, {Ely}, {Jenness}, {Labrie}, {Lim}, {Pierfederici}, {Pontzen},
  {Ptak}, {Refsdal}, {Servillat}, \& {Streicher}}]{astropy}
{Astropy Collaboration}, {Robitaille}, T.~P., {Tollerud}, E.~J., {et~al.} 2013,
  Astronomy and Astrophysics, 558, A33

\bibitem[{{Barnes}(2003)}]{barnes2003}
{Barnes}, S.~A. 2003, apj, 586, 464

\bibitem[{{Barrado y Navascues} \& {Stauffer}(1996)}]{barradoynavascues1996}
{Barrado y Navascues}, D., \& {Stauffer}, J.~R. 1996, Astronomy and
  Astrophysics, 310, 879

\bibitem[{{Bilir} {et~al.}(2008){Bilir}, {Ak}, {Karaali}, {Cabrera-Lavers},
  {Chonis}, \& {Gaskell}}]{bilir2008}
{Bilir}, S., {Ak}, S., {Karaali}, S., {et~al.} 2008, mnras, 384, 1178

\bibitem[{{Brandt} \& {Huang}(2015{\natexlab{a}})}]{brandt2015-1}
{Brandt}, T.~D., \& {Huang}, C.~X. 2015{\natexlab{a}}, The Astrophysical
  Journal, 807, 24

\bibitem[{{Brandt} \& {Huang}(2015{\natexlab{b}})}]{brandt2015}
---. 2015{\natexlab{b}}, The Astrophysical Journal, 807, 58

\bibitem[{{Castellani} {et~al.}(2001){Castellani}, {Degl'Innocenti}, \& {Prada
  Moroni}}]{castellani2001}
{Castellani}, V., {Degl'Innocenti}, S., \& {Prada Moroni}, P.~G. 2001, Monthly
  Notices of the Royal Astronomical Society, 320, 66

\bibitem[{{Cochran} {et~al.}(2002){Cochran}, {Hatzes}, \&
  {Paulson}}]{cochran2002}
{Cochran}, W.~D., {Hatzes}, A.~P., \& {Paulson}, D.~B. 2002, The Astronomical
  Journal, 124, 565

\bibitem[{{Copenhagen University} {et~al.}(2006){Copenhagen University},
  {Institute}, {Cambridge}, {Uk}, \& {Real Instituto Y Observatorio de La
  Armada}}]{cmc14}
{Copenhagen University}, O., {Institute}, A.~O., {Cambridge}, {Uk}, \& {Real
  Instituto Y Observatorio de La Armada}, F. E.~S. 2006, VizieR Online Data
  Catalog, 1304, 0

\bibitem[{{Curtis} {et~al.}(2013){Curtis}, {Wolfgang}, {Wright}, {Brewer}, \&
  {Johnson}}]{curtis2013}
{Curtis}, J.~L., {Wolfgang}, A., {Wright}, J.~T., {Brewer}, J.~M., \&
  {Johnson}, J.~A. 2013, The Astronomical Journal, 145, 134

\bibitem[{{David} {et~al.}(2016){David}, {Conroy}, {Hillenbrand}, {Stassun},
  {Stauffer}, {Rebull}, {Cody}, {Isaacson}, {Howard}, \& {Aigrain}}]{david2016}
{David}, T.~J., {Conroy}, K.~E., {Hillenbrand}, L.~A., {et~al.} 2016,
  arXiv:1602.01901 {[}astro-ph]

\bibitem[{{Delorme} {et~al.}(2011){Delorme}, {Collier Cameron}, {Hebb},
  {Rostron}, {Lister}, {Norton}, {Pollacco}, \& {West}}]{delorme2011}
{Delorme}, P., {Collier Cameron}, A., {Hebb}, L., {et~al.} 2011, mnras, 413,
  2218

\bibitem[{{Donati}(2011)}]{donati2011}
{Donati}, J.-F. 2011, in , 23--31

\bibitem[{{Donati} \& {Landstreet}(2009)}]{donati2009}
{Donati}, J.-F., \& {Landstreet}, J.~D. 2009, Annual Review of Astronomy and
  Astrophysics, 47, 333

\bibitem[{{Douglas} {et~al.}(2014){Douglas}, {Ag{\"u}eros}, {Covey}, {Bowsher},
  {Bochanski}, {Cargile}, {Kraus}, {Law}, {Lemonias}, {Arce}, {Fierroz}, \&
  {Kundert}}]{douglas2014}
{Douglas}, S.~T., {Ag{\"u}eros}, M.~A., {Covey}, K.~R., {et~al.} 2014, apj,
  795, 161

\bibitem[{{Duch{\^e}ne} {et~al.}(2013){Duch{\^e}ne}, {Bouvier}, {Moraux},
  {Bouy}, {Konopacky}, \& {Ghez}}]{duchene2013}
{Duch{\^e}ne}, G., {Bouvier}, J., {Moraux}, E., {et~al.} 2013, Astronomy and
  Astrophysics, 555, A137

\bibitem[{{Friedman}(1984)}]{friedman1984}
{Friedman}, J.~H. 1984, A variable span scatterplot smoother

\bibitem[{{Garraffo} {et~al.}(2015{\natexlab{a}}){Garraffo}, {Drake}, \&
  {Cohen}}]{garraffo2015}
{Garraffo}, C., {Drake}, J.~J., \& {Cohen}, O. 2015{\natexlab{a}}, The
  Astrophysical Journal, 813, 40

\bibitem[{{Garraffo} {et~al.}(2015{\natexlab{b}}){Garraffo}, {Drake}, \&
  {Cohen}}]{garraffo2015-1}
---. 2015{\natexlab{b}}, The Astrophysical Journal Letters, 807, L6

\bibitem[{{Gieren} {et~al.}(1998){Gieren}, {Fouqu{\'e}}, \&
  {G{\'o}mez}}]{gieren1998}
{Gieren}, W.~P., {Fouqu{\'e}}, P., \& {G{\'o}mez}, M. 1998, The Astrophysical
  Journal, 496, 17

\bibitem[{{Gizis} \& {Reid}(1995)}]{gizis1995}
{Gizis}, J., \& {Reid}, I.~N. 1995, The Astronomical Journal, 110, 1248

\bibitem[{{Goldman} {et~al.}(2013){Goldman}, {R{\"o}ser}, {Schilbach},
  {Magnier}, {Olczak}, {Henning}, {Juri{\'c}}, {Schlafly}, {Chen}, {Platais},
  {Burgett}, {Hodapp}, {Heasley}, {Kudritzki}, {Morgan}, {Price}, {Tonry}, \&
  {Wainscoat}}]{goldman2013}
{Goldman}, B., {R{\"o}ser}, S., {Schilbach}, E., {et~al.} 2013, aap, 559, A43

\bibitem[{{Griffin}(2012)}]{griffin2012}
{Griffin}, R.~F. 2012, Journal of Astrophysics and Astronomy, 33, 29

\bibitem[{{Griffin} {et~al.}(1988){Griffin}, {Griffin}, {Gunn}, \&
  {Zimmerman}}]{griffin1988}
{Griffin}, R.~F., {Griffin}, R. E.~M., {Gunn}, J.~E., \& {Zimmerman}, B.~A.
  1988, The Astronomical Journal, 96, 172

\bibitem[{{Griffin} {et~al.}(1982){Griffin}, {Mayor}, \& {Gunn}}]{griffin1982}
{Griffin}, R.~F., {Mayor}, M., \& {Gunn}, J.~E. 1982, Astronomy and
  Astrophysics, 106, 221

\bibitem[{{Guenther} {et~al.}(2005){Guenther}, {Paulson}, {Cochran},
  {Patience}, {Hatzes}, \& {Macintosh}}]{guenther2005}
{Guenther}, E.~W., {Paulson}, D.~B., {Cochran}, W.~D., {et~al.} 2005, Astronomy
  and Astrophysics, 442, 1031

\bibitem[{{Hartman} {et~al.}(2011){Hartman}, {Bakos}, {Noyes}, {Sip{\H o}cz},
  {Kov{\'a}cs}, {Mazeh}, {Shporer}, \& {P{\'a}l}}]{hartman2011}
{Hartman}, J.~D., {Bakos}, G.~{\'A}., {Noyes}, R.~W., {et~al.} 2011, The
  Astronomical Journal, 141, 166

\bibitem[{{Henden} {et~al.}(2012){Henden}, {Levine}, {Terrell}, {Smith}, \&
  {Welch}}]{apass}
{Henden}, A.~A., {Levine}, S.~E., {Terrell}, D., {Smith}, T.~C., \& {Welch}, D.
  2012, Journal of the American Association of Variable Star Observers
  (JAAVSO), 40, 430

\bibitem[{{Hodgkin} {et~al.}(1999){Hodgkin}, {Pinfield}, {Jameson}, {Steele},
  {Cossburn}, \& {Hambly}}]{hodgkin1999}
{Hodgkin}, S.~T., {Pinfield}, D.~J., {Jameson}, R.~F., {et~al.} 1999, mnras,
  310, 87

\bibitem[{{Howell}(2006)}]{howell2006}
{Howell}, S.~B. 2006, Handbook of {{CCD Astronomy}}, 2nd edn., Cambridge
  Observing Handbooks for Research Astronomers ({Cambridge University Press})

\bibitem[{{Howell} {et~al.}(2014){Howell}, {Sobeck}, {Haas}, {Still},
  {Barclay}, {Mullally}, {Troeltzsch}, {Aigrain}, {Bryson}, {Caldwell},
  {Chaplin}, {Cochran}, {Huber}, {Marcy}, {Miglio}, {Najita}, {Smith},
  {Twicken}, \& {Fortney}}]{howell2014}
{Howell}, S.~B., {Sobeck}, C., {Haas}, M., {et~al.} 2014, Publications of the
  Astronomical Society of the Pacific, 126, 398

\bibitem[{{Irwin} \& {Bouvier}(2009)}]{irwin2009}
{Irwin}, J., \& {Bouvier}, J. 2009, in IAU Symposium, Vol. 258, {{IAU
  Symposium}}, ed. {E. E. Mamajek, D. R. Soderblom, \& R. F. G. Wyse}, 363--374

\bibitem[{{Janson} {et~al.}(2014){Janson}, {Bergfors}, {Brandner},
  {Kudryavtseva}, {Hormuth}, {Hippler}, \& {Henning}}]{janson2014}
{Janson}, M., {Bergfors}, C., {Brandner}, W., {et~al.} 2014, The Astrophysical
  Journal, 789, 102

\bibitem[{{Jester} {et~al.}(2005){Jester}, {Schneider}, {Richards}, {Green},
  {Schmidt}, {Hall}, {Strauss}, {Vanden Berk}, {Stoughton}, {Gunn},
  {Brinkmann}, {Kent}, {Smith}, {Tucker}, \& {Yanny}}]{jester2005}
{Jester}, S., {Schneider}, D.~P., {Richards}, G.~T., {et~al.} 2005, The
  Astronomical Journal, 130, 873

\bibitem[{{Kaiser} {et~al.}(2002){Kaiser}, {Aussel}, {Burke}, {Boesgaard},
  {Chambers}, {Chun}, {Heasley}, {Hodapp}, {Hunt}, {Jedicke}, {Jewitt},
  {Kudritzki}, {Luppino}, {Maberry}, {Magnier}, {Monet}, {Onaka}, {Pickles},
  {Rhoads}, {Simon}, {Szalay}, {Szapudi}, {Tholen}, {Tonry}, {Waterson}, \&
  {Wick}}]{panstarrs}
{Kaiser}, N., {Aussel}, H., {Burke}, B.~E., {et~al.} 2002, in , 154--164

\bibitem[{{Kopytova} {et~al.}(2016){Kopytova}, {Brandner}, {Tognelli}, {Prada
  Moroni}, {Da Rio}, {R{\"o}ser}, \& {Schilbach}}]{kopytova2016}
{Kopytova}, T.~G., {Brandner}, W., {Tognelli}, E., {et~al.} 2016, Astronomy and
  Astrophysics, 585, A7

\bibitem[{{Kraus} {et~al.}(2015){Kraus}, {Cody}, {Covey}, {Rizzuto}, {Mann}, \&
  {Ireland}}]{kraus2015}
{Kraus}, A.~L., {Cody}, A.~M., {Covey}, K.~R., {et~al.} 2015, The Astrophysical
  Journal, 807, 3

\bibitem[{{Kraus} \& {Hillenbrand}(2007)}]{adam2007}
{Kraus}, A.~L., \& {Hillenbrand}, L.~A. 2007, aj, 134, 2340

\bibitem[{{Mann} {et~al.}(2015){Mann}, {Gaidos}, {Mace}, {Johnson}, {Bowler},
  {LaCourse}, {Jacobs}, {Vanderburg}, {Kraus}, {Kaplan}, \& {Jaffe}}]{mann2015}
{Mann}, A.~W., {Gaidos}, E., {Mace}, G.~N., {et~al.} 2015, arXiv:1512.00483
  {[}astro-ph]

\bibitem[{{Mason} {et~al.}(2001){Mason}, {Wycoff}, {Hartkopf}, {Douglass}, \&
  {Worley}}]{mason2001}
{Mason}, B.~D., {Wycoff}, G.~L., {Hartkopf}, W.~I., {Douglass}, G.~G., \&
  {Worley}, C.~E. 2001, The Astronomical Journal, 122, 3466

\bibitem[{{Matsunaga} {et~al.}(2009){Matsunaga}, {Feast}, \&
  {Menzies}}]{matsunaga2009}
{Matsunaga}, N., {Feast}, M.~W., \& {Menzies}, J.~W. 2009, Monthly Notices of
  the Royal Astronomical Society, 397, 933

\bibitem[{{Matt} {et~al.}(2015){Matt}, {Brun}, {Baraffe}, {Bouvier}, \&
  {Chabrier}}]{matt2015}
{Matt}, S.~P., {Brun}, A.~S., {Baraffe}, I., {Bouvier}, J., \& {Chabrier}, G.
  2015, The Astrophysical Journal Letters, 799, L23

\bibitem[{{McQuillan} {et~al.}(2013){McQuillan}, {Aigrain}, \&
  {Mazeh}}]{mcquillan2013}
{McQuillan}, A., {Aigrain}, S., \& {Mazeh}, T. 2013, mnras, 432, 1203

\bibitem[{{Mermilliod} {et~al.}(2009){Mermilliod}, {Mayor}, \&
  {Udry}}]{mermilliod2009}
{Mermilliod}, J.-C., {Mayor}, M., \& {Udry}, S. 2009, Astronomy and
  Astrophysics, 498, 949

\bibitem[{{Morgan} {et~al.}(2012){Morgan}, {West}, {Garc{\'e}s}, {Catal{\'a}n},
  {Dhital}, {Fuchs}, \& {Silvestri}}]{morgan2012}
{Morgan}, D.~P., {West}, A.~A., {Garc{\'e}s}, A., {et~al.} 2012, The
  Astronomical Journal, 144, 93

\bibitem[{{Morzinski}(2011)}]{morzinski2011}
{Morzinski}, K.~M. 2011, Ph.D. Thesis, 180

\bibitem[{{Newton} {et~al.}(2015){Newton}, {Irwin}, {Charbonneau},
  {Berta-Thompson}, {Dittmann}, \& {West}}]{newton2015}
{Newton}, E.~R., {Irwin}, J., {Charbonneau}, D., {et~al.} 2015,
  arXiv:1511.00957 {[}astro-ph]

\bibitem[{{Ochsenbein} {et~al.}(2000){Ochsenbein}, {Bauer}, \&
  {Marcout}}]{Ochsenbein2000}
{Ochsenbein}, F., {Bauer}, P., \& {Marcout}, J. 2000, aaps, 143, 23

\bibitem[{{Patience} {et~al.}(1998){Patience}, {Ghez}, {Reid}, {Weinberger}, \&
  {Matthews}}]{patience1998}
{Patience}, J., {Ghez}, A.~M., {Reid}, I.~N., {Weinberger}, A.~J., \&
  {Matthews}, K. 1998, The Astronomical Journal, 115, 1972

\bibitem[{{Paulson} {et~al.}(2004){Paulson}, {Cochran}, \&
  {Hatzes}}]{paulson2004}
{Paulson}, D.~B., {Cochran}, W.~D., \& {Hatzes}, A.~P. 2004, The Astronomical
  Journal, 127, 3579

\bibitem[{{Paulson} {et~al.}(2003){Paulson}, {Sneden}, \&
  {Cochran}}]{paulson2003}
{Paulson}, D.~B., {Sneden}, C., \& {Cochran}, W.~D. 2003, The Astronomical
  Journal, 125, 3185

\bibitem[{{Perryman} {et~al.}(1997){Perryman}, {Lindegren}, {Kovalevsky},
  {Hoeg}, {Bastian}, {Bernacca}, {Cr{\'e}z{\'e}}, {Donati}, {Grenon},
  {Grewing}, {van Leeuwen}, {van der Marel}, {Mignard}, {Murray}, {Le Poole},
  {Schrijver}, {Turon}, {Arenou}, {Froeschl{\'e}}, \& {Petersen}}]{hip}
{Perryman}, M. A.~C., {Lindegren}, L., {Kovalevsky}, J., {et~al.} 1997, aap,
  323, L49

\bibitem[{{Perryman} {et~al.}(1998){Perryman}, {Brown}, {Lebreton}, {Gomez},
  {Turon}, {Cayrel de Strobel}, {Mermilliod}, {Robichon}, {Kovalevsky}, \&
  {Crifo}}]{perryman1998}
{Perryman}, M. A.~C., {Brown}, A. G.~A., {Lebreton}, Y., {et~al.} 1998,
  Astronomy and Astrophysics, 331, 81

\bibitem[{{Pojma{\'n}ski}(2002)}]{ASAS}
{Pojma{\'n}ski}, G. 2002, Acta Astronomica, 52, 397

\bibitem[{{Poppenhaeger} \& {Wolk}(2014)}]{katja2014}
{Poppenhaeger}, K., \& {Wolk}, S.~J. 2014, aap, 565, L1

\bibitem[{{Pourbaix} {et~al.}(2004){Pourbaix}, {Tokovinin}, {Batten}, {Fekel},
  {Hartkopf}, {Levato}, {Morrell}, {Torres}, \& {Udry}}]{pourbaix2004}
{Pourbaix}, D., {Tokovinin}, A.~A., {Batten}, A.~H., {et~al.} 2004, Astronomy
  and Astrophysics, 424, 727

\bibitem[{{Press} \& {Rybicki}(1989)}]{press1989}
{Press}, W.~H., \& {Rybicki}, G.~B. 1989, The Astrophysical Journal, 338, 277

\bibitem[{{Prosser} {et~al.}(1995){Prosser}, {Shetrone}, {Dasgupta}, {Backman},
  {Laaksonen}, {Baker}, {Marschall}, {Whitney}, {Kuijken}, \&
  {Stauffer}}]{prosser1995}
{Prosser}, C.~F., {Shetrone}, M.~D., {Dasgupta}, A., {et~al.} 1995,
  Publications of the Astronomical Society of the Pacific, 107, 211

\bibitem[{{Quinn} {et~al.}(2014){Quinn}, {White}, {Latham}, {Buchhave},
  {Torres}, {Stefanik}, {Berlind}, {Bieryla}, {Calkins}, {Esquerdo}, {F{\H
  u}r{\'e}sz}, {Geary}, \& {Szentgyorgyi}}]{quinn2014}
{Quinn}, S.~N., {White}, R.~J., {Latham}, D.~W., {et~al.} 2014, The
  Astrophysical Journal, 787, 27

\bibitem[{{Radick} {et~al.}(1995){Radick}, {Lockwood}, {Skiff}, \&
  {Thompson}}]{radick1995}
{Radick}, R.~R., {Lockwood}, G.~W., {Skiff}, B.~A., \& {Thompson}, D.~T. 1995,
  apj, 452, 332

\bibitem[{{Radick} {et~al.}(1987){Radick}, {Thompson}, {Lockwood}, {Duncan}, \&
  {Baggett}}]{radick1987}
{Radick}, R.~R., {Thompson}, D.~T., {Lockwood}, G.~W., {Duncan}, D.~K., \&
  {Baggett}, W.~E. 1987, apj, 321, 459

\bibitem[{{Reid} \& {Gizis}(1997)}]{reid1997}
{Reid}, I.~N., \& {Gizis}, J.~E. 1997, The Astronomical Journal, 114, 1992

\bibitem[{{Reid} \& {Mahoney}(2000)}]{reid2000}
{Reid}, I.~N., \& {Mahoney}, S. 2000, Monthly Notices of the Royal Astronomical
  Society, 316, 827

\bibitem[{{Reid}(1992)}]{reid1992}
{Reid}, N. 1992, Monthly Notices of the Royal Astronomical Society, 257, 257

\bibitem[{{Reiners} \& {Mohanty}(2012)}]{reiners2012}
{Reiners}, A., \& {Mohanty}, S. 2012, apj, 746, 43

\bibitem[{{Richichi} \& {Percheron}(2002)}]{richichi2002}
{Richichi}, A., \& {Percheron}, I. 2002, Astronomy and Astrophysics, 386, 492

\bibitem[{{R{\"o}ser} {et~al.}(2011){R{\"o}ser}, {Schilbach}, {Piskunov},
  {Kharchenko}, \& {Scholz}}]{roser2011}
{R{\"o}ser}, S., {Schilbach}, E., {Piskunov}, A.~E., {Kharchenko}, N.~V., \&
  {Scholz}, R.-D. 2011, aap, 531, A92

\bibitem[{{Skrutskie} {et~al.}(2006){Skrutskie}, {Cutri}, {Stiening},
  {Weinberg}, {Schneider}, {Carpenter}, {Beichman}, {Capps}, {Chester},
  {Elias}, {Huchra}, {Liebert}, {Lonsdale}, {Monet}, {Price}, {Seitzer},
  {Jarrett}, {Kirkpatrick}, {Gizis}, {Howard}, {Evans}, {Fowler}, {Fullmer},
  {Hurt}, {Light}, {Kopan}, {Marsh}, {McCallon}, {Tam}, {Van Dyk}, \&
  {Wheelock}}]{2mass}
{Skrutskie}, M.~F., {Cutri}, R.~M., {Stiening}, R., {et~al.} 2006, aj, 131,
  1163

\bibitem[{{Skumanich}(1972)}]{skumanich72}
{Skumanich}, A. 1972, apj, 171, 565

\bibitem[{{Steele} \& {Jameson}(1995)}]{steele1995}
{Steele}, I.~A., \& {Jameson}, R.~F. 1995, mnras, 272, 630

\bibitem[{{Tokovinin}(1997)}]{tokovinin1997}
{Tokovinin}, A.~A. 1997, Astronomy and Astrophysics Supplement Series, 124, 75

\bibitem[{{Tonry} {et~al.}(2012){Tonry}, {Stubbs}, {Lykke}, {Doherty},
  {Shivvers}, {Burgett}, {Chambers}, {Hodapp}, {Kaiser}, {Kudritzki},
  {Magnier}, {Morgan}, {Price}, \& {Wainscoat}}]{tonry2012}
{Tonry}, J.~L., {Stubbs}, C.~W., {Lykke}, K.~R., {et~al.} 2012, The
  Astrophysical Journal, 750, 99

\bibitem[{{van Leeuwen}(2009)}]{vanleeuwen2009}
{van Leeuwen}, F. 2009, aap, 497, 209

\bibitem[{{Vanderburg} \& {Johnson}(2014)}]{vanderburg2014}
{Vanderburg}, A., \& {Johnson}, J.~A. 2014, Publications of the Astronomical
  Society of the Pacific, 126, 948

\bibitem[{{Walkowicz} {et~al.}(2013){Walkowicz}, {Basri}, \&
  {Valenti}}]{walkowicz2013}
{Walkowicz}, L.~M., {Basri}, G., \& {Valenti}, J.~A. 2013, The Astrophysical
  Journal Supplement Series, 205, 17

\bibitem[{{Wright} {et~al.}(2010){Wright}, {Eisenhardt}, {Mainzer}, {Ressler},
  {Cutri}, {Jarrett}, {Kirkpatrick}, {Padgett}, {McMillan}, {Skrutskie},
  {Stanford}, {Cohen}, {Walker}, {Mather}, {Leisawitz}, {Gautier}, {McLean},
  {Benford}, {Lonsdale}, {Blain}, {Mendez}, {Irace}, {Duval}, {Liu}, {Royer},
  {Heinrichsen}, {Howard}, {Shannon}, {Kendall}, {Walsh}, {Larsen}, {Cardon},
  {Schick}, {Schwalm}, {Abid}, {Fabinsky}, {Naes}, \& {Tsai}}]{wise}
{Wright}, E.~L., {Eisenhardt}, P. R.~M., {Mainzer}, A.~K., {et~al.} 2010, The
  Astronomical Journal, 140, 1868

\bibitem[{{Zacharias} {et~al.}(2012){Zacharias}, {Finch}, {Girard}, {Henden},
  {Bartlett}, {Monet}, \& {Zacharias}}]{zacharias2012}
{Zacharias}, N., {Finch}, C.~T., {Girard}, T.~M., {et~al.} 2012, VizieR Online
  Data Catalog, 1322, 0

\end{thebibliography}
\setlength{\baselineskip}{1.667\baselineskip}

\clearpage
\pagebreak
\clearpage
\begin{turnpage}
\begin{deluxetable*}{ccccccrcccccl}[p]
\tablewidth{0pt}
\tabletypesize{\scriptsize}
\tablecaption{Rotation periods measured from {\it K2} data \label{tab:k2}}
\tablehead{
\colhead{Ros.}& \colhead{}& \colhead{}& \colhead{}& \colhead{$K_p$}& \colhead{Prev.}& \colhead{{\it K2}}& \colhead{Peak}& \colhead{}& \colhead{}& \colhead{Spot}& \colhead{Multi}& \colhead{}\\ 
 \colhead{No.\tablenotemark{a}}& \colhead{EPIC}& \colhead{RA}& \colhead{Dec}& \colhead{Mag}& \colhead{$P_{rot}$\tablenotemark{b}}& \colhead{$P_{rot}$ (d)}& \colhead{Power}& \colhead{Q\tablenotemark{c}}& \colhead{Bin\tablenotemark{d}}& \colhead{Evol.\tablenotemark{e}}& \colhead{$P_{rot}$\tablenotemark{f}}& \colhead{notes}
}
\startdata
64 & 210942999 & 03:34:10.57 & +22:04:21.17 & 13.7 & -& 1.81 & 0.86 & 0& 2 & N & N& \nodata\\
78 & 210865372 & 03:43:47.09 & +20:51:36.25 & 13.4 & D& 12.32 & 0.63 & 0& 0 & Y & N& \nodata\\
80 & 210499476 & 03:46:59.57 & +15:24:16.93 & 14.2 & -& 0.86 & 0.75 & 0& 2 & N & N& not quite sinusoidal pattern\\
85 & 211058178 & 03:51:03.11 & +23:54:13.14 & 9.8 & A& 12.57 & 0.56 & 0& 2 & Y & N& amplitude drops to almost 0 as the campaign progresses\\
86 & 211169647 & 03:52:41.02 & +25:48:16.21 & 10.6 & -& 14.66 & 0.63 & 0& 0 & Y & M& \nodata\\
87 & 211137806 & 03:52:42.71 & +25:12:26.03 & 15.4 & -& 14.38 & 0.72 & 0& 0 & Y & N& \nodata\\
98 & 210894955 & 03:55:36.90 & +21:18:48.29 & 14.6 & -& 0.73 & 0.09 & 1& 0 & N & N& using secondary $P_{rot}$\\
99 & 210704853 & 03:55:57.18 & +18:25:56.31 & 13.1 & -& 23.12 & 0.62 & 1& 0 & N & Y& periodic signal includes large dips; possible eclipse signature? \\
104 & 210365286 & 03:58:14.38 & +12:37:40.88 & 14.1 & -& 0.87 & 0.31 & 0& 0 & N & N& \nodata\\
108 & 211201094 & 03:59:09.65 & +26:28:34.43 & 10.9 & -& 15.26 & 0.85 & 0& 0 & N & N& \nodata\\
109 & 210941195 & 03:59:14.19 & +22:02:37.86 & 13.4 & -& 20.73 & 0.61 & 1& 0 & M & M& periodic signal includes large dips; possible eclipse signature? \\
111 & 210579563 & 03:59:15.03 & +16:39:51.94 & 14.5 & -& 1.28 & 0.79 & 0& 0 & N & N& \nodata\\
113 & 211026112 & 03:59:26.62 & +23:23:07.59 & 13.8 & -& 1.32 & 0.66 & 0& 0 & N & N& \nodata\\
116 & 210769813 & 04:00:15.59 & +19:24:36.47 & 13.9 & -& 1.86 & 0.92 & 0& 0 & N & N& \nodata\\
117 & 210419589 & 04:00:44.96 & +13:54:22.75 & 13.5 & -& 2.66 & 0.54 & 0& 2 & N & Y& two clear periods\\
125 & 210772796 & 04:03:39.04 & +19:27:18.03 & 9.9 & D& 11.60 & 0.81 & 0& 2 & Y & N& \nodata\\
127 & 210523892 & 04:04:10.74 & +15:48:21.90 & 16.6 & -& 0.30 & 0.27 & 1& 2 & M & N& not quite sinusoidal pattern\\
128 & 210742592 & 04:04:12.81 & +18:59:44.60 & 15.3 & -& 0.80 & 0.34 & 0& 0 & Y & N& \nodata\\
133 & 210771915 & 04:05:25.67 & +19:26:31.78 & 10.9 & DA& 13.85 & 0.60 & 0& 1 & Y & N& \nodata\\
138 & 210744677 & 04:06:20.65 & +19:01:39.03 & 13.6 & -& 0.48 & 0.59 & 1& 0 & N & N& \nodata\\
148 & 211049709 & 04:08:36.21 & +23:46:07.01 & 10.3 & D& 9.39 & 0.96 & 0& 0 & Y & N& \nodata\\
149 & 211036776 & 04:08:40.18 & +23:33:25.74 & 12.8 & -& 21.02 & 0.90 & 0& 2 & Y & N& \nodata\\
\nodata & 210371851 & 04:11:06.42 & +12:47:48.17 & 14.9 & -& 0.90 & 0.16 & 1& 0 & M & Y& using secondary $P_{rot}$\\
161 & 211041649 & 04:11:56.22 & +23:38:10.77 & 9.2 & -& 2.31 & 0.59 & 0& 2 & Y & N& \nodata\\
166 & 210963067 & 04:12:47.36 & +22:23:26.96 & 14.7 & -& 1.84 & 0.57 & 0& 0 & N & N& extracted using {\it daofind}\\
169 & 210490365 & 04:13:05.62 & +15:14:51.95 & 14.5 & -& 1.89 & 0.10 & 1& 2 & Y & N& using secondary $P_{rot}$; not all instrumental noise removed\\
170 & 210497173 & 04:13:52.37 & +15:21:54.06 & 13.7 & -& 1.85 & 0.69 & 0& 2 & M & Y& flares?\\
197 & 210432080 & 04:16:25.42 & +14:10:17.12 & 14.1 & -& 1.45 & 0.28 & 1& 2 & Y & M& using secondary $P_{rot}$; not all instrumental noise removed\\
\nodata & 210980637 & 04:17:06.88 & +22:40:23.93 & 9.4 & -& 49.32 & 0.98 & 0& 2 & M & N& unusually long for this age\\
202 & 210744818 & 04:17:25.15 & +19:01:47.67 & 10.4 & D& 12.84 & 0.46 & 1& 0 & Y & N& not all instrumental noise removed\\
203 & 210470484 & 04:17:28.14 & +14:54:03.62 & 13.3 & D& 2.38 & 0.44 & 1& 0 & Y & N& not all instrumental noise removed\\
207 & 210408563 & 04:17:47.69 & +13:39:42.36 & 11.8 & -& 8.69 & 0.48 & 1& 2 & Y & Y& possibly a half-period harmonic; additional $\approx$1~d period visible\\
209 & 210489127 & 04:17:51.62 & +15:13:37.66 & 14.4 & -& 3.65 & 0.67 & 0& 0 & N & N& \nodata\\
210 & 210451321 & 04:17:55.57 & +14:32:46.47 & 13.9 & -& 10.12 & 0.87 & 0& 0 & Y & N& \nodata\\
214 & 211019716 & 04:18:10.79 & +23:17:04.75 & 10.0 & -& 1.86 & 0.72 & 0& 2 & M & N& amplitude increases over campaign\\
220 & 210700098 & 04:18:33.85 & +18:21:52.91 & 13.8 & -& 2.34 & 0.88 & 0& 0 & N & N& \nodata\\
235 & 210736105 & 04:20:27.63 & +18:53:49.84 & 13.3 & -& 20.31 & 0.67 & 1& 0 & N & N& angular peak\\
\nodata & 210489654 & 04:20:47.99 & +15:14:09.07 & 12.6 & -& 43.59 & 0.84 & 0& 2 & N & N& unusually long for this age\\
247 & 210459199 & 04:21:35.10 & +14:41:42.82 & 14.4 & -& 3.64 & 0.45 & 0& 0 & N & N& not all instrumental noise removed\\
\nodata & 210774807 & 04:21:49.56 & +19:29:08.68 & 14.1 & -& 0.20 & 0.19 & 1& 2 & Y & Y& two peaks have almost same height, not harmonics\\
267 & 210786154 & 04:23:22.85 & +19:39:31.23 & 9.2 & DA& 10.02 & 0.92 & 0& 2 & M & N& \nodata\\
277 & 210718930 & 04:23:56.68 & +18:38:20.15 & 14.6 & -& 2.42 & 0.29 & 1& 0 & N & N& not all instrumental noise removed\\
288 & 210674406 & 04:24:16.94 & +18:00:10.47 & 9.8 & DA& 11.47 & 0.64 & 0& 0 & Y & N& \nodata\\
303 & 210741091 & 04:25:14.56 & +18:58:24.89 & 12.0 & D& 10.97 & 0.91 & 0& 2 & Y & N& \nodata\\
316 & 210640966 & 04:25:47.02 & +17:32:40.58 & 14.6 & -& 2.56 & 0.31 & 1& 2 & N & N& not all instrumental noise removed\\
328 & 210674207 & 04:26:21.71 & +18:00:01.06 & 14.6 & -& 1.05 & 0.31 & 1& 2 & N & Y& possibly another $\approx$1~d period not detected in secondary light curve\\
347 & 210709514 & 04:27:58.96 & +18:30:00.88 & 9.8 & -& 11.13 & 0.56 & 0& 2 & Y & Y& \nodata\\
355 & 210651981 & 04:28:28.78 & +17:41:45.39 & 11.2 & PD& 2.44 & 0.71 & 0& 2 & Y & N& blended neighbor plus a possibly spurious {\it daofind} source\enddata
\tablecomments{This table is available in its entirety in a machine-readable form in the online journal. A portion is shown here for guidance regarding its form and content.}
\tablenotetext{a}{\citet{roser2011}}
\tablenotetext{b}{Gives source of previous $P_{rot}$ measurement, if any. P is \citet{prosser1995}, D is \citet{delorme2011}, and A is from our analysis of ASAS data (Cargile et al. in prep). There is no overlap between the \citet{radick1987, radick1995}, \citet{hartman2011} and {\it K2} samples.}
\tablenotetext{c}{Quality flag. 0 is a confident detection, 1 indicates residual noise or confusion in the periodogram.}
\tablenotetext{d}{Indicated binary status (Section \ref{binaries}, Table \ref{tab:bin}). 0 indicates a single star (to the best of our knowledge), 1 is a photometrically identified candidate binary, and 2 is a confirmed binary from the literature.}
\tablenotetext{e}{Indicates visible spot evolution over several cycles.  ``Y,'' ``M,'' and ``N'' flags indicate ``Yes,'' ``Maybe,'' and ``No.''}
\tablenotetext{f}{Indicates multiple periods in the periodogram.  ``Y,'' ``M,'' and ``N'' flags indicate ``Yes,'' ``Maybe,'' and ``No.''}
\end{deluxetable*}
\end{turnpage}\clearpage
\global\pdfpageattr\expandafter{\the\pdfpageattr/Rotate 90}






\end{document}